\documentclass[11pt,letter]{article}
\usepackage[utf8]{inputenc}
\usepackage[T1]{fontenc}
\usepackage{amsmath}
\usepackage{amsfonts}
\usepackage{amssymb}
\usepackage{booktabs}
\usepackage{graphicx}
\usepackage{float}
\usepackage[usenames,dvipsnames]{xcolor}
\usepackage{colortbl}
\usepackage[titletoc,toc]{appendix}
\usepackage{empheq}
\usepackage{slashed}
\usepackage{cancel}
\usepackage{bbm}
\usepackage{amsthm}
\usepackage{braket}
\usepackage{tcolorbox}
\usepackage{upgreek}
\usepackage{relsize}
\usepackage{simplewick}
\usepackage{multirow}
\usepackage{todonotes}
\usepackage{mathrsfs} 
\usepackage{dsfont}
\usepackage{jheppub}
\usepackage{yhmath}
\usepackage{tikz}
\usepackage[normalem]{ulem}
\usepackage{orcidlink}
\usetikzlibrary{calc,matrix,decorations.pathmorphing,decorations.markings,arrows,positioning,intersections,mindmap,backgrounds,arrows.meta,decorations.pathreplacing}
\usepackage[export]{adjustbox}

\usepackage{datetime}

\usepackage{tikz}
\usetikzlibrary{shapes,arrows.meta,decorations.pathmorphing,backgrounds,positioning,fit,petri}
\definecolor{charcoal}{HTML}{343837}

\definecolor{darkyellow}{rgb}{0.5, 0.5, 0.0}
\definecolor{darkpurple}{rgb}{0.5, 0.2, 0.8}
\definecolor{darkblue}{rgb}{0.0, 0.0, 0.8}
\definecolor{darkgreen}{rgb}{0.0, 0.4, 0.0}
\definecolor{darkred}{rgb}{0.5, 0.0, 0.0}

\usepackage{hyperref}
\hypersetup{
    linktocpage,
     colorlinks,
     citecolor=darkgreen,
     linkcolor= darkgreen,
     urlcolor=darkgreen
}

\def \d {\mathrm{d}}

\newcommand{\la}{\langle}
\newcommand{\ra}{\rangle}

\newcommand{\bs}[1]{ {\boldsymbol{#1}} }

\newcommand{\df}{\mathrm{d}}

\newcommand{\nn}{\nonumber}

\newcommand{\ep}{\epsilon}

\renewcommand{\-}{{\dt-}}

\allowdisplaybreaks[4]

\title{Toward the Analytic Bootstrap of Energy Correlators}

\author[a,b]{Jianyu Gong\orcidlink{0000-0002-3325-5777},}
\emailAdd{jianyu\_gong@sjtu.edu.cn}

\author[c]{Andrzej Pokraka\orcidlink{0000-0003-1186-4624},}
\emailAdd{apokraka@uva.nl}

\author[a,b]{Kai Yan\orcidlink{0000-0001-8327-7061},}
\emailAdd{yan.kai@sjtu.edu.cn}

\author[d,e]{Xiaoyuan Zhang\orcidlink{0000-0002-2090-2381}}
\emailAdd{xyz2@mit.edu}

\affiliation[a]{State Key Laboratory of Dark Matter Physics, Shanghai Key Laboratory for Particle Physics and Cosmology, Key Laboratory for Particle Astrophysics and Cosmology (MOE), School of Physics and Astronomy, Shanghai Jiao Tong University, Shanghai 200240, China}
\affiliation[b]{Institute of Nuclear and Particle Physics (INPAC), Shanghai Jiao Tong University, Shanghai 200240, China}
\affiliation[c]{%
    Institute of Physics, University of Amsterdam, Amsterdam, 1098 XH, The Netherlands
}
\affiliation[d]{Department of Physics, Harvard University, Cambridge, MA 02138, USA}
\affiliation[e]{Center for Theoretical Physics -- a Leinweber Institute, Massachusetts Institute of Technology, Cambridge, MA 02139, USA}

\preprint{MIT-CTP/5912}

\abstract{
	In this paper, we present a framework for the analytic bootstrap of three-point energy correlators, a crucial observable in $\mathcal{N}=4$ super Yang-Mills theory and quantum chromodynamics (QCD). Our approach combines spherical contour techniques, general physical constraints such as pole cancellations, and power correction data in the singular limits to determine its analytic expression. In contrast to previous bootstrap studies restricted to scattering amplitudes for supersymmetric theories, our framework makes use of the properties of Feynman integrals, marking a significant step toward bootstrapping realistic QCD observables. Using this method, we derive analytic expressions for leading-order three-point energy correlators with equal and unequal energy weights, where the latter are crucial ingredients for projected $N$-point energy correlators. We also apply the recently developed technique of analytic regression with lattice reduction as a way to bypass needing explicit expressions for the singular limits.
    Bridging theoretical advances in scattering amplitudes with the renewed interest in weighted cross-sections, our work opens the door to precision tests of QCD dynamics through analytic event-shape predictions.
}

\begin{document}

\maketitle
\flushbottom
\newpage

\section{Introduction}

In recent years, the energy correlator has become a popular observable in both collider physics and the formal studies of quantum field theory. Defined as an energy-weighted cross section, energy correlators measure the energy deposited in the detectors as a function of the angles between any pair of the detectors. With energy weights, they are insensitive to soft gluon divergence and serve as natural observables for studying strong interactions. In the early days of quantum chromodynamics (QCD), the two-point energy correlator was used to prove asymptotic freedom. More recently, advances in calculation techniques of both perturbative QCD and conformal field theories have enabled precise theoretical predictions for energy correlators and access to both perturbative and non-perturbative physics.

The two-point energy correlator, originally called energy-energy correlation function (EEC), was proposed in the 1970s~\cite{Basham:1978zq,Basham:1978bw}, as an event shape observable at $e^+e^-$ colliders. In perturbation theory, the EEC is defined as 
\begin{equation}
    \frac{\d \sigma}{\d x}\equiv\sum_{i,j}\int \d\sigma \times\frac{E_i E_j}{Q^2}\times\delta\left(x-\frac{1-\cos\theta_{ij}}{2}\right)\,,
\end{equation}
where $\d\sigma$ is the differential cross-section of the scattering process, $E_i$ is the energy of the particle $i$, $\theta_{ij}$ is the angle between particle $i$ and $j$, and $Q$ is the total energy. The sum $i,j$ runs over all final-state particles. The analytic results for the EEC have been known to next-to-next-to-leading order (NNLO) for $\mathcal{N}=4$ super Yang-Mills (SYM)~\cite{Belitsky:2013xxa,Belitsky:2013bja,Belitsky:2013ofa,Henn:2019gkr} and to NLO for QCD and Higgs decays~\cite{Dixon:2018qgp,Luo:2019nig,Gao:2020vyx}. 
Subsequently, the EEC has been generalized to a broader family of observables called $N$-point energy correlators~\cite{Chen:2019bpb},
\begin{equation}\label{eq:EC_def}
    \frac{\df\sigma}{\df x_{12}\cdots \df x_{(N-1)N}} \equiv \sum_{m}\sum_{1\leq i_1,\cdots i_N \leq m}\int\! \df \sigma_{m}\times  \prod_{1\leq k \leq N}\frac{E_{i_k}}{Q} \prod_{1\leq j < l \leq N}\delta\left(x_{jl}-\frac{1-\cos\theta_{i_{j}i_{l}}}{2}\right)
    \,.
\end{equation}
Here $m$ is the number of final-state particles and $\df\sigma_m$ is the differential cross section for $m$ final-state particles. The sum over $i_1,\cdots i_N$ runs over all $N$ subsets of the final states. The infrared safety is guaranteed by summing over different final states that include both virtual corrections and real emissions. 
The LO three-point correlator (EEEC)
has been computed in $\mathcal{N}=4$ SYM, QCD and Higgs decays~\cite{Yan:2022cye,Yang:2022tgm,Yang:2024gcn}. 
Very recently, the collinear limit of four-point correlator (EEEEC) was computed in $\mathcal{N}=4$ SYM~\cite{Chicherin:2024ifn}. 
The integrands for the $N$-point energy correlators have also inspired a more systematic study of the module square of the scattering amplitudes~\cite{He:2024hbb,Bourjaily:2025iad} and form factors~\cite{He:2025zbz}, allowing the first studies on the analytic phase-space integrations.

The analytic results for the EEEC and EEEEC results are expressed in terms of transcendental functions. 
Most are polylogarithms, but elliptic polylogarithms start to show up in the EEC at NNLO and the EEEEC at LO with arbitrary angles, as explored in Ref.~\cite{Henn:2019gkr,Ma:2025qtx}. 
The scale dependence in $N$-point energy correlators is rather rich since there are $N(N-1)/2$ angular variables. 
Of particular interest are several kinematic limits such as collinear limits (particles moving almost along the same direction, either homogeneous or with hierarchies) and the coplanar limit (falling onto the same plane).
A summary of kinematic limits in the three-point correlator is given in Ref.~\cite{Yang:2022tgm}.

Precision calculations of energy correlators has enabled phenomenological studies of the Standard Model in collider experiments. 
They can be interpreted either as event shapes at $e^+e^-$ colliders, or as jet and jet substructure observables at hadron colliders. 
Experimentally, energy correlators have been measured precisely in recent years and applied to many precision studies of high-energy collider physics, including QCD scatterings~\cite{Dixon:2019uzg,
Chen:2020vvp,Lee:2022uwt,Chen:2023zlx,Gao:2019ojf,Gao:2023ivm,Gao:2024wcg,
COLLINS1981381,ELLIS198499,deFlorian:2004mp,Tulipant:2017ybb,Moult:2018jzp,Moult:2019vou,Ebert:2020sfi,Moult:2022xzt,Duhr:2022yyp}, top mass measurement~\cite{Holguin:2022epo,Holguin:2024tkz,Xiao:2024rol}, electroweak physics~\cite{Holguin:2023bjf,Alipour-fard:2025dvp} and Higgs decays~\cite{Gao:2020vyx,Lin:2024lsj}. 
In particular, an effective field theory approach is applied to resum the large logarithms in the collinear limit for the three-point correlator~\cite{Chen:2023zlx} and the analytic result leads to the currently most precise $\alpha_s$ extraction in the jet substructure~\cite{CMS:2024mlf}.
Additionally, there is also growing interest in using energy correlators in heavy-ion colliders to probe QCD dynamics at lower energies~\cite{Andres:2022ovj,Yang:2023dwc,Bossi:2024qho}.

Importantly, energy correlators are well-defined observables both perturbatively and non-perturbatively, and can offer new insights into the dynamics of quantum field theory.
Perturbatively, the availability of analytic results has revealed hidden structures and properties similar to scattering amplitudes. 
Expressed in terms of integrated amplitudes (or squared form factor), energy correlators share both common and distinct features compared to standard Feynman integrals. 
Through the three-point and four-point analytic results, we can explore the symbol alphabets, symmetry groups in the function space, as well as other analytic properties. 
For example, it is found in Ref.~\cite{Yan:2022cye} that EEEC kinematic space in $\mathcal{N}=4$ 
can be embedded into a hexagon on a unit circle and the remaining symmetry is characterized by a dihedral group $D_6$. This
implies that EEEC contains some similar structures to the six-gluon scattering amplitude. Such properties are also preserved in QCD, since at the integrand level, it shares a similar form to $\mathcal{N}=4$. 
This similarity, suggests the possibility of building a framework for bootstrapping physical observables. 
Non-perturbatively, energy correlators can also be defined in terms of correlation functions of light-transformed local operators~\cite{Hofman:2008ar}.
Such observables probe the properties of states that are created due to localized excitations in the field theory and exhibit an OPE in the multi-collinear limit~\cite{Kravchuk:2018htv,Kologlu:2019mfz}. 
In Refs.~\cite{Chen:2020adz, Chen:2021gdk, Chang:2022ryc, Chen:2022jhb}, the light-ray OPE analysis is used to study the power corrections in the collinear limit of energy correlators and, more recently, an OPE-based framework is developed to study non-perturbative QCD corrections~\cite{Lee:2024esz, Chen:2024nyc, Lee:2025okn, Chang:2025kgq, Guo:2025zwb, Herrmann:2025fqy, Kang:2025zto}.
In order to deepen our understanding of quantum field theories and energy correlators, advanced theoretical techniques are required and more analytic data is needed.

In this paper, we focus on a class of finite $N$-point energy correlator integrals. From Eq.~\eqref{eq:EC_def}, we note that the differential cross-section receives contributions from both finite terms $i_1\neq i_2\cdots \neq i_N$ and the so-called contact terms, where some of the $i_k$ are the same. In these scenarios, some of the delta functions vanish and the powers of energy weights become more than one. In other words, the energy correlators can be recast in terms of the following integrals
\begin{equation}\label{eq:ECrecast_def}
    F(\{x_{jl}\}, \{a_k\}) \equiv \sum_{m}\sum_{1\leq i_1 < \cdots < i_N \leq m}\int\! \df \sigma_{m}\times  \prod_{1 \leq k \leq N}\frac{E_{i_k}^{a_k}}{Q} \prod_{1\leq j < l \leq N}\delta\left(x_{jl}-\frac{1-\cos\theta_{i_{j}i_{l}}}{2}\right)
    \,.
\end{equation}
In general, these powers $a_k$ can be zero; however, in this work, we will only study $a_k>0$ cases, such that there is no soft divergence. Such integrals are also useful for evaluating the projection of multi-point correlators. In Ref.~\cite{Chen:2020vvp}, the authors proposed the projected energy correlators that only measure the maximal angular distance $x_L=\max\{x_{jl}\}$, and this observable has been widely used in many precision QCD studies at the LHC (e.g., see~\cite{Lee:2022uwt,Chen:2023zlx}). In fixed-order calculation, projected energy correlators are naturally written in terms of these $F$ integrals (c.f. Eq.~(45) of Ref.~\cite{Chen:2020vvp}). Therefore, obtaining their analytic form will significantly advance the studies of projected energy correlators and their phenomenological applications. 
While direct calculation is possible for these integrals, in this work, we exploit the analytic properties of the integrand and develop a bootstrap framework to obtain their analytic expressions, avoiding tedious integrations and simplification. 
We also hope that the methods applied here will generalize and consequently streamline the computation of higher-loop and higher-point energy correlators.

The current difficulty of obtaining analytic expressions in high-energy physics has led to a resurgent interest in exploring various bootstrap methods. 
To the best of our knowledge, there are mainly two types of perturbative bootstrap programs: amplitude bootstrap and Landau bootstrap. The {\it amplitude bootstrap}~\cite{Dixon:2016nkn,Caron-Huot:2016owq,Caron-Huot:2019vjl,Dixon:2020bbt,Dixon:2022rse,Basso:2024hlx,Cai:2024znx,Cai:2025atc,Guo:2021bym,Dersy:2022bym} works on the level of amplitudes and makes use of their properties, such as unitarity, gauge invariance, supersymmetry and so on. 
This allows for the determination of the ``DNA'' of amplitudes --- the {\it Symbol}~\cite{Goncharov:2010jf,dna} --- which can be lifted to transcendental functions. 
For example, the authors of Ref.~\cite{Dixon:2020bbt, Dixon:2022rse} bootstrap the three-point form factor in planar $\mathcal{N}=4$ to eight-loop at the amplitude level. Despite lots of progress in $\mathcal{N}=4$, generalizing the amplitude bootstrap programs to QCD is still challenging: (i) the amplitude is no longer uniform transcendental; (ii) one needs to determine rational functions instead of rational numbers. Moreover, individual rational structures contain spurious higher-degree poles that conspire to cancel in the physical results, leading to additional complexity in the leading singularities.
On the other hand, the {\it Landau bootstrap} (also called Feynman integral bootstrap) \cite{Chicherin:2017dob,Caron-Huot:2018dsv,Henn:2018cdp,He:2021fwf,He:2021eec,Morales:2022csr} works at the level of individual Feynman diagrams (or Feynman integrals) using only the analytic properties of these integrals. 
Although there are not many physical constraints in individual diagrams, one can use tools like Landau analysis~\cite{Bourjaily:2020wvq,Hannesdottir:2024cnn,Hannesdottir:2024hke} to access the singularities and symbol alphabets. 
In this paper, we use elements of both bootstrap frameworks to obtain analytic expressions for energy correlators in both $\mathcal{N}=4$ and QCD.

As a first step, we present the bootstrap method for three-point energy correlators at LO ($N=3$ and $m=4$ in Eq.~\eqref{eq:ECrecast_def}).
For simplicity, we will focus on the homogeneous collinear limit, where all angles $\theta_{kl}$ are sent to zero at the same time. In the collinear limit, the energy correlator integrals simplify and share forms similar to the Feynman parameter integrals. 
Our strategy is to obtain an initial {\it input} by computing all leading singularities at the leading transcendental weight of the integrated splitting functions 
and then to fix the lower-weight remainders from physical constraints and OPE limits at leading and subleading powers.
The outline of this paper is as follows. In Sec.~\ref{sec:preliminaries}, we set up the integrals in the collinear limit and introduce the spherical contour methods~\cite{Arkani-Hamed:2017ahv}. In particular, we present the algorithm for computing spherical contours and illustrate it with several examples. In Sec.~\ref{sec:bootstrap3}, we discuss the bootstrap method for obtaining the lower-weight expressions for energy correlators. In Sec.~\ref{sec:bootstrap-lattice}, we demonstrate an alternative approach, lattice reduction, to replace some steps of the bootstrap workflow. We apply this alternative method to obtain the analytic result for several higher energy weights. We conclude in Sec.~\ref{sec:conclusion}.

\section{Preliminaries}\label{sec:preliminaries}

\subsection{Collinear limit of three-point correlator}
\label{sec:factorization}
In this subsection, we set up the explicit integrals that we want to study with the bootstrap method. Following the definition in Eq.~\eqref{eq:ECrecast_def}, the full three-point correlator contains three angle measurements, and at leading order, the nontrivial distribution involves four final-state particles:
\begin{align}
\label{eq:eeeclo_def}
    &\text{EEEC}(x_{12},x_{13},x_{23};a,b,c)\nn\\
    =& \sum_{i, j, k}
  \int \d \sigma \frac{E_i^a E_j^b E_k^c}{Q^{a+b+c}} \delta \left( x_{12} -
  \frac{1 - \cos \theta_{12}}{2} \right) \delta \left( x_{13} - \frac{1 -
  \cos \theta_{13}}{2} \right) \delta \left( x_{23} - \frac{1 - \cos
  \theta_{23}}{2} \right)\nn\\
   =& \sum_{i, j, k} \int \d \text{PS}_4 |
  \mathcal{M}^{\text{tree}}_{1\rightarrow 4} |^2 \frac{E_i^a E_j^b E_k^c}{Q^{a+b+c}}\,\delta \left( \zeta_{12} -
  \frac{1 - \cos \theta_{12}}{2} \right) \delta \left( x_{13} - \frac{1 -
  \cos \theta_{13}}{2} \right) \nn\\
  &\times \delta \left( x_{23} - \frac{1 - \cos
  \theta_{23}}{2} \right)+ \mathcal{O}(\alpha_s^3)\,.
\end{align}
Here $\d\text{PS}_4$ is the four-particle phase space and  $|\mathcal{M}^{\text{tree}}_{1 \rightarrow 4} |^2$ is the tree-level $1\to 4$ matrix elements. In QCD, $|\mathcal{M}^{\text{tree}}_{1 \rightarrow 4} |^2$ is the tree-level amplitude square of $\gamma^\star \to 4$ partons for $e^+e^-\rightarrow$ hadrons process, and is the tree-level amplitude square of $\text{Higgs}\to 4$ partons for hadronic Higgs decays. In $\mathcal{N}=4$ SYM, $|\mathcal{M}^{\text{tree}}_{1 \rightarrow 4} |^2$ is the super form factor square $\left|\mathcal{F}_4\right|^2$ created by a half-BPS operator.

In the triple collinear limit, $x_{12}\sim x_{13}\sim x_{23}\to 0$, both the phase space and matrix elements at this order, can be factored:
\begin{equation}\begin{aligned}
  \text{dPS}_4 &\approx \text{dPS}_2 \times \d \sigma_3^{\text{coll}}
  \,, 
  \\
  |
  \mathcal{M}^{\text{tree}}_{1 \rightarrow 4} |^2 &\approx |
  \mathcal{M}^{\text{tree}}_{1 \rightarrow 2} |^2 \times  |
  \mathcal{M}_{1 \rightarrow 3}^{\text{coll}, \text{tree}} |^2
  \,, 
  \\
  |
  \mathcal{M}_{1 \rightarrow 3}^{\text{coll}, \text{tree}} |^2
  &= \left(
  \frac{\mu^2 e^{\gamma_E}}{4 \pi} \right)^{2 \epsilon} \frac{4
  g^4}{s_{123}^2} \times P_{1 \rightarrow 3}^{\text{tree}}
  \,,
\end{aligned} \end{equation}
where $\d\sigma_3^{\text{coll}}$ is the 3-particle collinear phase space and
$| \mathcal{M}_{1 \rightarrow 3}^{\text{coll}} |^2$ is the collinear $1
\rightarrow 3$ matrix element, which can be expressed in terms of splitting
function $P_{1 \rightarrow 3}^{\text{tree}}$. Substituting them into the LO EEEC in Eq.~\eqref{eq:eeeclo_def}, yields the LO collinear EEEC: 
\begin{align}
\label{eq:EEEC_fac}
  \text{EEEC}(x_{12},x_{13},x_{23};a,b,c) \approx \sigma_{0} 
   \sum_{i, j, k} \int \!\d \sigma_3^{\text{coll}}  |
  \mathcal{M}_{1 \rightarrow 3}^{\text{coll}, \text{tree}} |^2 \frac{E_i^a E_j^b E_k^c}{Q^{a+b+c}} \delta^3 \!\left( x_{i j} {-} \frac{1 {-} \cos \theta_{i j}}{2}
  \right).
\end{align}
Here, the normalization $\sigma_{0}= \int \text{dPS}_2  | \mathcal{M}^{\text{tree}}_{1\rightarrow 2}|^2$ is the Born total cross-section of the given process. Note that here $\delta^3 \left[ x_{i j} - (1 - \cos \theta_{i j})/2\right]$ is the short-hand notation for the three angle measurements in Eq.~\eqref{eq:eeeclo_def}.

The triple collinear phase space can be expressed in terms of the Lorentz invariant quantities $s_{ij}$ and the energy fraction $z_i$ of the final-state partons~\cite{Ritzmann:2014mka}:
\begin{equation}
  \d \sigma_3^{\text{coll}} = \text{ds}_{12} \text{ds}_{13} \text{ds}_{23}
  \text{dz}_1 \text{dz}_2 \text{dz}_3 \delta (1 - z_1 - z_2 - z_3) \frac{4
  \Theta (- \Delta_3^{\text{coll}}) (- \Delta_3^{\text{coll}})^{- \frac{1}{2}
  - \epsilon}}{(4 \pi)^{5 - 2 \epsilon} \Gamma (1 - 2 \epsilon)}\,,
\end{equation}
where
\begin{equation}
  \Delta_3^{\text{coll}} = (z_3 s_{12} - z_1 s_{23} - z_2 s_{13})^2 - 4 z_1 z_2 s_{13} s_{23}\,.
\end{equation}
is the collinear version of the Gram determinant. 
Parameterizing the Lorentz invariants in terms of angles $s_{ij}=z_i z_j (1-\cos\theta_{ij})/2$, the $\text{ds}_{ij}$ integration over the $\delta$ functions becomes straightforward. 
Moreover, with this parameterization, the dependence on the integration variables in the collinear Gram determinant factors out
\begin{equation}
    \Delta_3^{\text{coll}}=(z_1 z_2 z_3)^2 \tilde{\Delta}_3,\quad \tilde{\Delta}_3=x_{12}^2+x_{13}^2+x_{23}^2-2x_{12}x_{13}-2x_{13}x_{23}-2x_{12}x_{23}\,,
\end{equation}
where $\tilde{\Delta}_3$ is a constant.

Regarding the collinear matrix elements, as discussed above, we only need the $\mathcal{N}=4$, quark and gluon $1\rightarrow 3$ tree-level splitting functions for LO collinear EEEC. 
The $\mathcal{N}=4$ splitting function is 
\begin{equation}
\label{eq:N4_split}
    P_{1\to 3}(z_1, z_2, z_3) 
    {=} N_c^2 \left[ \frac{s_{123}^2}{2 s_{13} s_{23}} \!\left( \frac{1}{z_1 z_2} {+} \frac{1}{(1-z_1)(1-z_2)} \right)\! {+} \frac{s_{123}}{s_{12} z_3} \!\left( \frac{1}{z_1} {+} \frac{1}{1-z_1} \right)\! {+} \text{perms} \right]\!,
\end{equation}
where $N_c$ is the number of colors and ``perms'' denotes the permutation of three final-state particles. The quark jet and gluon jet contain different partonic processes with different color factors~\cite{Campbell:1997hg,Catani:1998nv,Ritzmann:2014mka}:
\begin{align}\label{eq:qcdchannel}
    P_{q\rightarrow 3}(z_1, z_2, z_3)&=C_F T_F n_f\times P_{q q^\prime \bar q^\prime} + C_F (C_A-2C_F)\times P_{q q\bar q}+C_F^2\times P_{qgg}^{(C_F)}+C_F C_A\times P_{qgg}^{(C_A)}\,,\nn\\
    P_{g\rightarrow 3}(z_1,z_2,z_3) &= C_F T_F n_f \times P_{gq\bar q}^{(C_F)}+C_A T_F n_f \times P_{gq\bar q}^{(C_A)}+ C_A^2 \times P_{ggg}\,.
\end{align}
Note that the subscripts on the right-hand side represent the explicit final-state particles after splitting. Their analytic expressions are summarized in App.~\ref{app:splitting_function}.

Additionally, the collinear three-point correlator contains a $S_3$ symmetry, arising from the permutation of three angles $(x_{12},x_{13},x_{23})$.
Therefore, we only need to analyze one measurement term, e.g. $i=1,j=2,k=3$ in Eq.~\eqref{eq:EEEC_fac}, and
all others can be recovered from the $i=1,j=2,k=3$ term. 
Putting this all together yields the following expression for LO collinear EEEC:
\begin{align}
    \frac{\d^3\sigma_{a,b,c}}{\d x_{12} \d x_{13} \d x_{23}}&=\left(
  \frac{\mu^2 e^{\gamma_E}}{4 \pi} \right)^{2 \epsilon}\frac{2^{4-a-b-c} g^4 \Theta(-\tilde{\Delta}_3)}{(4\pi)^{5-2\epsilon} \Gamma(1-2\epsilon)} (-\tilde{\Delta}_3)^{-\frac{1}{2}-\epsilon}\nn\\
  &\times \int \d z_1 \d z_2 \d z_3\, \delta(1-z_1-z_2-z_3) z_1^{a+1-2\epsilon} z_2^{b+1-2\epsilon} z_3^{c+1-2\epsilon}\times \frac{P_{1\to 3}(z_1,z_2,z_3)}{s_{123}^2}\nn\\
  &+\text{perms of }(x_{12},\, x_{13},\,x_{23})\,,
\end{align}
where the $S_3$ symmetry is encoded in the permutations of $(x_{12},x_{13},x_{23})$ on the last line. Note that this permutation is different from the permutation in Eq.~\eqref{eq:N4_split}, which comes from the splitting function itself. 

The square root $\sqrt{-\tilde \Delta_3}$ can be rationalized by the conformal ratio $\{z,\bar z\}$ that satisfies $z\bar z=x_{13}/x_{12}$ and $(1-z)(1-\bar z)=x_{23}/x_{12}$. For convenience, we can further set $x_{12}=1$ in the calculation and recover it by dimensional analysis in the end. 
This amounts to 
\begin{align}
    \frac{\d^3\sigma_{a,b,c}}{\d x_{12} \d z \d\bar z}&=\left(
  \frac{\mu^2 e^{\gamma_E}}{4 \pi} \right)^{2 \epsilon}\frac{2^{4-a-b-c} g^4 \Theta(-\tilde{\Delta}_3)}{(4\pi)^{5-2\epsilon} \Gamma(1-2\epsilon)} (-\tilde{\Delta}_3)^{-\frac{1}{2}-\epsilon}\times G(z)\,,
\end{align}
where $G(z)$ contains the sum over the six permutations after chaining the kinematic variables:
\begin{align}
\label{eq:permutations}
    G(z)&=
    G_0(z) + G_0(1 - z)\nn\\
    &+ \frac{1}{|1 - z|^4} \left( G_0\left( \frac{z}{z - 1} \right) + G_0\left( \frac{1}{1 - z} \right) \right)
    + \frac{1}{|z|^4} \left( G_0\left( \frac{1}{z} \right) + G_0\left( \frac{z - 1}{z} \right) \right)\,.
\end{align}
The first term has the integral representation
\begin{equation}
    G_0(z)=\int \d z_1 \d z_2 \d z_3\, \delta(1-z_1-z_2-z_3) z_1^{a+1-2\epsilon} z_2^{b+1-2\epsilon} z_3^{c+1-2\epsilon}\times \frac{P_{1\to 3}(z_1,z_2,z_3)}{s_{123}^2}\,,
    \label{eq:G0}
\end{equation}
where we omit the $a,b,c$ dependence in both $G(z)$ and $G_0(z)$. The $a=b=c=1$ case is the collinear LO EEEC computed in Ref.~\cite{Chen:2019bpb}, and the result has been used for both light-ray OPE analysis in quantum field theory~\cite{Chen:2020adz,Chen:2021gdk} and phenomenological analysis at colliders~\cite{Komiske:2022enw,Chen:2022swd}. The higher weights $a>1,b>1,c>1$ are not infrared-safe observables: while their LO results are finite in $d=4$ dimension, they will contain infrared divergences in higher-loop order. However, these infrared divergences can be absorbed by objects like fragmentation functions or track functions~\cite{Chang:2013rca,Chang:2013iba,Li:2021zcf,Jaarsma:2022kdd,Jaarsma:2023ell}, so that we can make reliable predictions for the higher energy weights and compare with experimental data. In the meantime, certain combinations of the higher energy weights will also contribute to the projected $N$-point correlators defined in Ref.~\cite{Chen:2020vvp}. For instance, the projected $4$-point receives contributions from $a+b+c=4$, namely $\{a,b,c\}=\{2,1,1\},\{1,2,1\}, \{1,1,2\}$. We will discuss the higher energy weight in detail in Sec.~\ref{sec:high-energy-weight}.

In the following, we will focus on developing a bootstrap method to obtain the analytic expression of $G(z)$ for both $\mathcal{N}=4$ SYM splittings and all partonic channels in QCD. For illustration, we will explain the case $a=b=c=1$ in detail.

\subsection{The spherical contour method}
\label{sec:spherical_contour}

In this section, we review the spherical contour method~\cite{Arkani-Hamed:2017ahv, YelleshpurSrikant:2019khx, Gong:2022erh}. 
We will use it to calculate the leading transcendentality of LO collinear EEEC $G(z)$.  
This method was originally developed to calculate a certain class of one-loop Feynman integrals in 4-dimensions, but can be extended to iterated integrals with quadric singularities (whose contour is a simplex). 
As shown in Eq.~\eqref{eq:G0}, the collinear EEEC is similar in form to one-loop Feynman integrals after introducing Feynman parameters to combine the denominators into a single denominator.
We explain the spherical contour algorithm, the implicit geometric configuration associated with the contour, and present the explicit calculation for some example integrals in the collinear EEEC.

Many calculations in quantum field theory involve Feynman integrals, and a large class of Feynman integrals can be expressed in terms of multiple polylogarithms (or called hyperlogarithms, generalized polylogarithms). 
Multiple polylogarithms are defined in terms of iterated integrals, and, in Ref.~\cite{Golden:2013xva}, the authors introduce a short-hand notation called Symbol~\cite{Goncharov:2010jf} (see also \cite{Duhr:2014woa} for more of a review article).  
The symbol of a polylogarithm conveniently forgets about most of the structure of the actual function, yet preserves all relations between polylogarithms: the symbol of a function and the function itself satisfy the same identities. 
The symbol also manifests the analytic structure of polylogarithmic functions through its relation to differentiation and taking discontinuities.

Each polylogarithmic function has an associated weight that, roughly speaking, corresponds to the minimal number of iterated integrals needed to represent the function.
A nice feature of polylogarithmic functions is that the derivative of a weight $n$ function $F_n$ is a linear combination of weight $(n{-}1)$ functions
\begin{equation}\label{eq:mpldiff}
    \mathrm{d}F_n=\sum_\alpha C_{\alpha}\; \mathrm{d}\log (s_\alpha) F_{\alpha,n-1}.
\end{equation}
One way to define the symbol of a weight $n$ polylogarithm function $\mathcal{S}F_n$ is through the above relation
\begin{equation}
    \mathcal{S}F_n=\sum_\alpha C_{\alpha}\; \mathcal{S}F_{\alpha,n-1}\otimes s_\alpha.
\end{equation}
Similarly, the derivative of the weight $(n-1)$ polylogarithms $F_{\alpha,n-1}$ is a linear combination of weight $(n-2)$ polylogarithms multiplied by a $\d\log$-form. 
Iterating the action of $\mathcal{S}$ the symbol of the weight $n$ polylogarithm $F_n$ is  
\begin{equation} \label{eq:Fnsymbol}
    \mathcal{S}F_n = \sum_{\alpha_i} C_{\alpha_1 \cdots \alpha_n}\; s_{\alpha_1} \otimes s_{\alpha_{2}} \otimes \cdots \otimes s_{\alpha_{n-1}} \otimes s_{\alpha_n}.
\end{equation} 
Up to integration constants, full function is schematically recovered from the symbol by 
\begin{align}
    F_n =\sum_{\alpha_i} 
    C_{\alpha_1\cdots\alpha_n}
    \int_{\gamma_n} \d\log s_{\alpha_n}(t_n) \cdots 
    \int_{\gamma_1} \d\log s_{\alpha_1}(t_1)
    + \text{integration constants}
    \,.
\end{align}
Here, we use $s_{\alpha_i}(t_i)$ to emphasize that $s_{\alpha_i}$ depends on the integration variable $t_i$ and the  $\gamma_i$ parametrize contours in the integration variable $t_i$.
See \cite{Duhr:2014woa} for more details on reconstructing the function from the symbol.

Since the derivative lowers the weight of $F_n$, the symbol of the derivative must have length $(n-1)$ instead of $n$.
From \eqref{eq:mpldiff}, it is clear that the symbol of  $\d F_n$ is obtained from \eqref{eq:Fnsymbol} by
chopping off the last entry of each term in the symbol:
\begin{equation}
    \mathcal{S}(\partial_{s_{\beta_n}} F_n) =\sum_{\alpha_i} \delta_{\alpha_n \beta_n}\;  C_{\alpha_1 \cdots \alpha_n}\; s_{\alpha_1}\otimes s_{\alpha_{2}}\otimes\cdots\otimes s_{\alpha_{n-1}}
    \,.
\end{equation}
On the other hand, the symbol of a discontinuity of $F_n$
is equivalent to chopping off the first entry of \eqref{eq:Fnsymbol}. 
Explicitly, the discontinuity of $F_n$ around the branch cut starting at $g_{\beta_1}=0$ and ending at $g_{\beta_1} = \infty$ has the symbol
\begin{equation}
    \mathcal{S}(\mathrm{Disc}_{s_{\beta_n}} F_n) = \sum_{\alpha_i} \delta_{\beta_1\alpha_1} \; 
    C_{\alpha_1 \cdots \alpha_n}\;
    s_{\alpha_{2}} \otimes \cdots \otimes s_{\alpha_{n-1}} \otimes s_{\alpha_n}.
\end{equation}
For an individual polylogarithm $F_n$, a spherical contour computes a discontinuity of $F_n$. 
Then, the symbol of $F_n$ can be obtained by repeated application of spherical contours. 

However, the integrals that appear in QCD EEEC computations are known to have mixed transcendental weight. 
Therefore, the maximal application of the spherical contour, which localizes the original integral, only produces the symbol for the highest weight polylogarithms contained in the expression of our integral; we will bootstrap the analytic form of the lower weight terms using the information from physical constraints and numeric evaluation.

\subsubsection{A first look at discontinuities }
\label{sec:logDisc}

In this section, we illustrate how to compute the discontinuities of polylogarithms from their integral expressions.
We start with a simple example: the logarithm; this can also be found in \cite{Arkani-Hamed:2012zlh}.
The logarithm has the following integral representation with a quadratic divisor
\begin{equation}\label{eq:example}
    \log(z)=\int_{0}^\infty\frac{(z-1)\mathrm{d}x_1}{(x_1+z)(x_1+1)}
    =\int_{0}^{\infty}\frac{(z-1) \mathrm{d}x_1\mathrm{d}x_2}{(x_1x_2+x_1+x_2+z)^2}
    =\int_{\Delta_3}\frac{\langle X\mathrm{d}X^2\rangle (LX)}{(XQX)^2}
    \,.
\end{equation}
The first and second integrals are defined in an affine space 
($x_1 \in \mathbb{C}$ and $(x_1, x_2) \in \mathbb{C}^2$) 
while the third is in projective space 
($X := [x_1: x_2 : x_3] \in )\,\,\mathbb{CP}^2$.%
\footnote{%
    Coordinate tuples in projective space $\mathbb{CP}^{n-1}$ are denoted with square brackets and colons instead of commas. 
    This is to remind us that two tuples are equivalent if they differ by an overall scaling: $[x_1:x_2,\dots,x_n] \simeq [\lambda x_1: \lambda x_1,\dots,\lambda x_n]$ for any $\lambda \in \mathbb{C}^*$. 
}
Here, 
\begin{equation}\label{eq:qtoyexample}
    L=[0:0:z-1]
    \qquad 
    \text{and}
    \qquad
    Q={\frac{1}{2}}\left(
    \begin{matrix}
        0&1&1\\
        1&0&1\\
        1&1&2z
    \end{matrix} 
    \right)\,,
\end{equation}
parameterize a line and a quadric hypersurface in the projective space. $(LX) = \sum_{i=1}^3 L_i X_i$ is the usual dot product and $(XQX) = \sum_{i,j=1}^3 X_i Q_{ij} X_j$. 
The angle brackets denote anti-symmetrization of indices and $\la X \d X^2 \ra$ is the usual volume form on 2-dimensional projective space. 
Explicitly, using indices and the Levi-Civita symbol, $\la X \d X^2 \ra = \sum_{i,j,k=1}^3 X_i\; \d X_j \wedge \d X_k\; \epsilon^{ijk}$. 

\begin{figure}
\begin{center}
    \includegraphics[scale=.5]{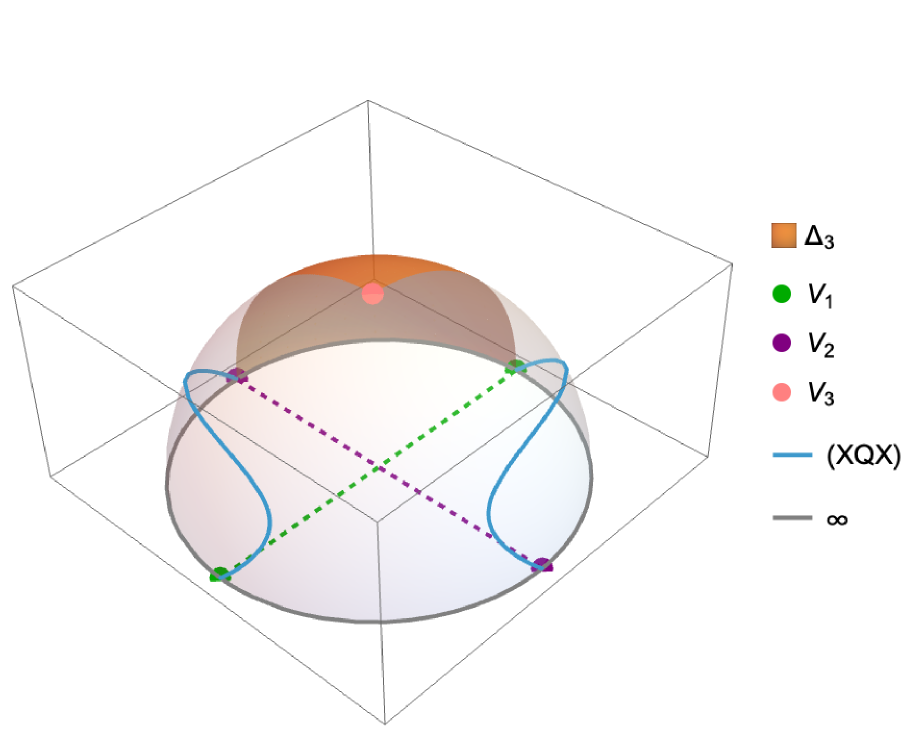}
    \caption{%
        A visualization of $\mathbb{CP}^2$ as a hemisphere. The line at infinity is placed at the equator and antipodal points are identified (as stressed by the dashed purple and green lines). 
        The standard simplex of $\Delta_3$ of $\mathbb{CP}^2$ is the area in orange and the curve $(XQX)=0$ is shown in blue. 
    }
    \label{fig:Delta3}
\end{center}
\end{figure}

Furthermore, $\Delta_3$ is the standard $2$-simplex of 2-dimensional complex projective space. 
It is the interior convex hull of the vertices $V_{i=1,\cdots,n=3}=[0:\cdots:1_i:\cdots:0]$ where $1_i$ denotes a one in the $i$th component: 
\begin{align}
    \Delta_3 = \{ X \in \mathbb{CP}^2 : X \in \mathrm{Conv}(V_1,\dots,V_3) \}
    \,.
\end{align}
To get some intuition for complex projective space, we can visualize the real slice of complex projective space as one hemisphere of the unit sphere (see Fig.~\ref{fig:Delta3}).
Here, the real points of $\Delta_3$, the surface $(XQX)=0$ and the line at infinity are shown on the hemisphere. 
Note that projective invariance requires the antipodal points on the line at infinity to be identified; this is why the vertices $V_1$ and $V_2$ appear twice and are connected by a colored dashed line.
Due to this identification, the quadric is continuous and not two disconnected pieces. 
Moreover, the vertex $V_3=[0:0:1]$ is the origin of usual affine space. 

To recover the affine integral from the projective integral, one can use the projective scale invariance to set any integration variable to $1$, i.e. $x_i=1$.
More generally, one can impose any linear condition on the integration variables to return to affine space (the Cheng-Wu theorem~\cite{Cheng:1987}). 

While the discontinuity of $\log(z)$ ($2\pi i$) is easily computed from the first representation in \eqref{eq:example},%
\footnote{%
    The standard way of computing this discontinuity is by understanding how the contour changes as $z$ loops around the origin in the first expression in \eqref{eq:example}. 
    As $z$ loops around the origin, the contour must move to avoid this pole
    \vspace{-1em}\begin{align*}
        \includegraphics[valign=c, scale=.9]{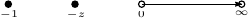}
        \quad\to\quad
        \includegraphics[valign=c, scale=.9]{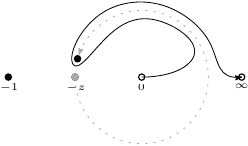}
        \quad\to\quad
        \includegraphics[valign=c, scale=.9]{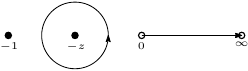}
        \,.
    \end{align*}
    The difference from the starting contour, otherwise known as the discontinuity, is simply the circular residue contour around the point $-z$. Taking the reside around $x_1=-z$ in \eqref{eq:example} yields $2\pi i$.
}
we want a contour that extracts this discontinuity from the last representation in \eqref{eq:example} with a quadratic divisor. 
For (pure) integrals with only quadric divisors $(XQX)$, there is an empirical conjecture that two integrations contribute to one weight. 
Additionally, the integration can be performed directly if there are at least two zeros on the diagonal of $Q$. 
In such cases, there exists a variable transformation that makes the quadric $(XQX)$ linear in a pair of integration variables that correspond to diagonal zeros of $Q$. 
For example, in the running example of the logarithm, the denominator is $(x_1+x_3)(x_2+x_3)+(z-1)x_3^2$. 
We can perform a transformation $y_+=x_1+x_3$, $y_-=x_2+x_3$ so that the denominator becomes linear in $y_\pm$: $(y_+y_-+(z-1)x_3^2)$. 

The problem now is: what contour in the variables $y_+$ and $y_-$ produces the discontinuities?
The answer is a two-dimensional contour called the \emph{Spherical Contour} $S^2$. 
This contour forces the two $y$'s into the same complex plane and then integrates over this whole plane. 
Explicitly, 
\begin{equation}
    S^2:= \{
        (y_+,y_-) \in \mathbb{C}^2 : 
        y_+=w,
        \quad 
        y_-=\bar{w},
        \quad 
        w \in \mathbb{C}
    \}.
\end{equation}
Since the $w$-complex plane can be compactified into a sphere, this is where the spherical contour $S^2$ gets its name. 
In order to perform the integration over the spherical contour, we make an additional variable transformation
\begin{equation}
    y_\pm=r \,e^{\pm i\phi},\quad r\in [0,\infty],\quad\phi\in[0,2\pi].
\end{equation}
Integration over this spherical contour yields the discontinuity 
\begin{equation}
    \int_{0}^\infty\mathrm{d}r \,(-2ir)\int_0^{2\pi}\mathrm{d}\phi\,\frac{(z-1)x_3^2}{(r^2+(z-1)x_3^2)^2}
    = - 2\pi i
    = - \mathrm{Disc}\log(z)
    .
\end{equation}
Here, we have set $x_3=1$ because the integral is projective and  the factor $-2ir$ comes from the volume-form on affine space
\begin{equation}
    \mathrm{d}x_1\wedge\mathrm{d}x_2 \mapsto(-2ir)\,\mathrm{d}r\wedge\mathrm{d}\phi.
\end{equation}
The minus sign in the last two equalities indicates that the real discontinuity contour has the opposite orientation to that of the spherical contour; the spherical contour must be compensated by an additional sign to extract the discontinuity.

A more involved example is the simplest weight-$2$ function: $\mathrm{Li}_2(z)$. 
It has an integral representation in the complex projective space ($X:=[x_1:\cdots:x_5] \in$) $\mathbb{CP}^4$ with a quadric divisor 
\begin{align}\label{eq:li2}
    \mathrm{Li}_2(z)&=\int_{\Delta_5}\frac{2z\,x_5\langle X\d X^4\rangle}{(x_5^2+x_5((1-z)x_1+x_2+x_3+x_4)+x_1x_3+x_1x_4+x_2x_4)^3},
    \nn\\
    &=\int_{\Delta_5}\frac{\la X\d X^4\ra(LX)}{(XQX)^3}
    :=\int_{\Delta_5}\la X\d X^4\ra\mathcal{I}(\mathrm{Li_2(z)})
    \,,
\end{align}
where 
\begin{equation}
    L=[0:0:0:0:2z], \quad \mathrm{and} \quad Q=\frac{1}{2}\left(\begin{matrix}
        0&0&1&1&1-z\\0&0&0&1&1\\1&0&0&0&1\\1&1&0&0&1\\1-z&1&1&1&2
    \end{matrix}\right).
\end{equation}
Here, $(LX)=0$ and $(XQX)=0$ define a linear and quadratic hypersurfaces in $\mathbb{CP}^5$ like for the previous example.

The symbol of $\mathrm{Li}_2(z)$ is known: $\mathcal{S}(\mathrm{Li}_2(z))=-((1-z)\otimes z)$ and its first entry%
\footnote{%
    Geometrically, the origin of this singularity comes from the $V_1V_5$ boundary of $\Delta_5$ intersecting the quadric. 
}
indicates that there is a branch cut in the complex $z$-plane whose endpoints are defined by $(1-z)^{\pm 1} =0$ (i.e., $z=1, \infty$). 
The corresponding discontinuity is $\mathrm{Disc}_{1-z}(\mathrm{Li}_2(z))=-2\pi i\log(z)$. 
To extract this discontinuity from the integral representation \eqref{eq:li2} using the spherical contour, we need a good choice first find a variable change that makes the quadratic divisor linear in two variables.%
\footnote{%
    Such a viable change exists for all quadrics in two or more variables.
}

Setting 
\begin{equation}
        x_1=\frac{(1-z)(y_1-y_5-x_2)+(1+z)(x_3+x_4)}{(1-z)^2},\quad
        x_5=\frac{(1-z)y_5-x_3-x_4}{1-z},
\end{equation}
makes the quadratic divisor linear in $y_1$ and $y_5$. 
Implementing this transformation yields the integrand 
\begin{equation}
    \mathcal{I}(\mathrm{Li_2(z)})=\frac{1}{1-z}\frac{2(1-z)^4\,z\,((1-z)y_5-x_3-x_4)}{((1-z)^2y_1y_5+z(x_3+x_4)^2-(1-z)x_2(x_3+zx_4))^3},
\end{equation}
where we have absorbed the Jacobian $1/(1-z)$ from the volume form. 
We also choose the kinematic region where $1-z>0$; 
one can obtain expressions in other regions by analytic continuation. 

In order to perform the spherical contour, we change variables again
\begin{equation}
    y_1=re^{-i\phi},~~y_5=re^{i\phi},\qquad r\in [0,\infty],~~\phi\in[0,2\pi]. 
\end{equation}
Then, the result of the spherical contour is 
\begin{align}
    &\int_{\Delta_3}\la X\d X^2\ra\int_0^\infty\d r\int_{0}^{2\pi}\d\phi\frac{2(-2ir)(1-z)^4(e^{i\phi}r(1-z)-x_3-x_4)}{(r^2(1-z)^2+z(x_3+x_4)^2-(1-z)x_2(x_3+zx_4))^3}\nn\\
    &=2\pi i\int_{\Delta_3}\la X\d X^2\ra\frac{(1-z)^2z(x_3+x_4)}{(z(x_3+x_4)^2-(1-z)x_2(x_3+zx_4))^2}\nn\\
    &=2\pi i\log(z)
    = -\mathrm{Disc}_{1-z}[\mathrm{Li}_2(z)],
\end{align}
which, up to a sign,
is the discontinuity of $\mathrm{Li}_2(z)$ across the branch cut between $z=1$ and $z=\infty$.
Again, we see that the spherical contour must be compensated by a minus sign to recover the desired discontinuity. 
This is because the spherical contour has the opposite orientation to the contour that extracts the discontinuity across the branch cut stretching from $z=1$ to $z=\infty$.

\subsubsection{The spherical contour algorithm}

In this section, we spell out general the algorithm for constructing the symbol of a polylogarithmic functions that have integral representations with quadratic divisors using spherical contours.
The general form of such an integral representation is
\begin{equation} \label{eq:Idef}
    I=\int_{\triangle}\frac{\langle X\mathrm{d}X^{n-1}\rangle\,T[X^k]}{(XQX)^{\frac{n+k}{2}}},
\end{equation}
where the projective measure is $\langle X\mathrm{d}X^{n-1}\rangle=\frac{1}{(n-1)!}\epsilon_{i_1\cdots i_n}X^{i_1}\mathrm{d}X^{i_2}\wedge\cdots\wedge\mathrm{d}X^{i_n}$, the Levi-Civita symbol is normalized by $\ep_{1\cdots n}=1$, the (tensor) numerator is $T[X^k]=T_{i_1\cdots i_k}X^{i_1}\cdots X^{i_k}$ and $\triangle$ is the standard $n$-simplex with has vertices $V_{i=1,\cdots,n+1}=[0:\cdots:1_i:\cdots:0]$.
Fortunately, the contours in energy correlator calculations are always the standard $n$-simplex.%
\footnote{%
    In general, any other $n$-simplex can be mapped into the standard $n$-simplex by a $\mathrm{PGL}(n)$ transformation. 
    Therefore, one only needs to consider the case of the standard simplex. 
}
Moreover, for \eqref{eq:Idef} to be a well defined projective integral, the integrand must be scale invariant; this restricts the exponent of $(XQX)$ to $(n+k)/2$.

The general form of the symbol for \eqref{eq:Idef} is
\begin{align}
    \mathcal{S}I 
    &= \sum_{w=1}^p
    \sum_{\bs{\alpha}_w } C_{\bs{\alpha}_w}\, s_{\alpha_1} \otimes \cdots \otimes s_{\alpha_w}
    \,,
    &
    p &:= \begin{cases}
        \frac{n}{2} & n \; \mathrm{even}
        \\
        \frac{n-1}{2} & n \; \mathrm{odd}
    \end{cases}
    \,,
\end{align}
where $p$ is the maximal weight of the polylogarithm $I$,  $s_{\alpha_i}$ are symbol entries, $C_{\bs{\alpha}_w}$ are constants, and $\bs{\alpha}_w$ is a multi-index with length $w$.
However, we will only use the spherical contour algorithm to compute the symbol of the maximal transcendental part of $I$: $I_\mathrm{max}$. 
The lower transcendental terms are bootstrapped in Sec.~\ref{sec:bootstrap3} and \ref{sec:bootstrap-lattice}. 
With this in mind, we focus only on the symbol of $I_\mathrm{max}$
\begin{align}
    \mathcal{S}I_\mathrm{max} 
    &= \sum_{\bs{\alpha}}
    C_{\bs{\alpha}}\, s_{\alpha_1} \otimes \cdots \otimes s_{\alpha_p}
    \,.
\end{align}
Its symbol is computed by taking all possible maximal sequential discontinuities via spherical contours.

The spherical contour method for computing discontinuities illustrated in Sec.~\ref{sec:logDisc}, naturally generalizes to integrals of the form Eq.~\eqref{eq:Idef}.
The discontinuities are computed by iterating the procedure of Sec.~\ref{sec:logDisc}; after each iteration one produces another integral of the form Eq.~\eqref{eq:Idef}.
Therefore, this procedure can be iterated until there are no more integrations left; the result of this maximally iterated discontinuity determines the coefficients $C_{\bs{\alpha}}$.
Moreover, each spherical contour is associated with a discontinuity $\mathrm{Disc}_{s_\bullet}$ and hence with a symbol entry.
Therefore, each non-trivial maximal sequence of spherical contours determines a sequence $s_{\alpha_1} \otimes \cdots \otimes s_{\alpha_n}$ in the symbol of $I_\mathrm{max}$. 

The spherical contour algorithm consists of three steps:
\vspace{-.5em}
\begin{itemize}
    \item [1.] {\bf{Determine a first symbol entry.}} 
        Choose two integration variables $x_i$ and $x_j$. 
        The corresponding spherical contour $S_{ij}^2$ computes the $\overline{ij}$-discontinuity of $I_\mathrm{max}$ (or one of its sequential discontinuities). 
        Here, $\overline{ij}=\overline{ji}$ is a polynomial in kinematic space that is the symbol entry related to the spherical contour in the variables $\{x_i,x_j\}$ (i.e., $\{\overline{ij}^{\pm} = 0\}$ are brach points of $I$ (or one of its sequential discontinuities)). 
        The explicit form of $\overline{ij}$ is given below.
        Note that if a choice of $(i,j)$ produces a symbol entry $\overline{ij}$ that is independent of kinematic variables, there is no discontinuity and one needs to make a different choice before moving to step 2.  
    
    \item [2.] {\bf{Compute discontinuities via spherical contour.}}
        Use the spherical contour to derive an integral representation for  the associated discontinuity.  
    \item [**] 
        Repeat steps 1 and 2 on the output of the previous step 2 until there is either no integrations remaining or only one integration remaining. 
        This determines a symbol term by tensoring all first entries from step 1:
        \begin{align}\begin{cases} \label{eq:symbolTerm}
            \overline{i_1 i_2} \otimes \cdots \otimes \overline{i_{n-1}i_{n}}
            =: s_{\alpha_1} \otimes \cdots \otimes s_{\alpha_{\frac{n}{2}}}
            & n \text{ even}
            \,,
            \\
            \overline{i_1 i_2} \otimes \cdots \otimes \overline{i_{n-2} i_{n-1}} =: s_{\alpha_1} \otimes \cdots \otimes s_{\alpha_{\frac{n-1}{2}}}
            & n \text{ odd}
            \,. 
        \end{cases}\end{align}
        Here, the ordered sequence of pairs $\{(i_1, i_2), \dots, (i_{n-1}, i_n)\}$ or $\{(i_1, i_2), \dots, (i_{n-2}, i_{n-1})\}$ records the choices made at each occurrence of step 1. 
        
    \item [3.] {\bf{Determine coefficients.}} 
        After maximal repetitions of steps 1 and 2, one is left with either a zero- or a one-fold integral that is trivial to evaluate due to projective invariance. 
        The coefficient of \eqref{eq:symbolTerm} is 
        \begin{align}\begin{aligned} \label{eq:coeffs}
             C_{i_1,i_2,\dots,i_{n-1},i_n}
             &= \left(\frac{1}{2 (2\pi i)}\right)^{\frac{n}{2}}
             \int_{S^2_{i_n i_{n-1}}} \cdots \int_{S^2_{i_1 i_2}} \mathcal{I}
             & n \text{ even}
             \,,
             \\
             C_{i_1,i_2,\dots,i_{n-1}}
             &= \left(\frac{1}{2 (2\pi i)}\right)^{\frac{n-1}{2}}
             \int_{S^2_{i_{n-1} i_{n-2}}} \cdots \int_{S^2_{i_1 i_2}} \mathcal{I}
             & n \text{ odd}
             \,,
        \end{aligned}\end{align}
        where $\mathcal{I}$ is the integrand of $I$.
        The powers  
        of $\frac{1}{2}$ are correlated to the specific representation of $\overline{ij}$. 
        We elaborate on this below. 
        
\end{itemize}
To determine every symbol term, the above steps need to be repeated for every possible sequence of pairs
\begin{equation}
    \mathcal{S}I_\mathrm{max}=\left\{
    \begin{aligned}
        &\sum_{\sigma \in S_n} C_{\sigma(1), \dots, \sigma(n)} \; \overline{\sigma(1)\sigma(2)}\otimes\cdots\otimes\overline{\sigma(n-1)\sigma(n)}
        && n~\mathrm{even},
        \\
        &\sum_{\sigma \in S_n} C_{\sigma(1), \dots, \sigma(n-1)} \; \overline{\sigma(1)\sigma(2)}\otimes\cdots\otimes\overline{\sigma(n-2)\sigma(n-1)}
        && n~\mathrm{odd},
    \end{aligned}
    \right. 
\end{equation}
where $S_n$ is the symmetric group of $n$ elements. 

In the following, we elaborate on the three steps involved in the spherical contour algorithm.
The discussion below assumes that it is the first application of the spherical contour $S_{ij}^2$ to $I$. 
If $m$ spherical contours have already been applied, one simply replaces $I \to \mathrm{Disc}_{\overline{i_{2m-1}i_{2m}}} \circ \cdots \circ \mathrm{Disc}_{\overline{i_1i_2}}[I]$.

\paragraph{The first entry.}
Here, we illustrate how to determine the first entry associated to the spherical contour $S_{ij}^2$.  
Choosing $x_i$ and $x_j$ as the active variables for the spherical contour (all others are held constant) also defines a $2\times2$ matrix $Q_{(ij)}$ that is simply the $i$'th and $j$'th rows/columns of the matrix $Q$
\begin{equation}
    Q_{(ij)}=\left(\begin{matrix}
        q_{ii}& q_{ij}\\
        q_{ij}& q_{jj}
    \end{matrix}\right)
    \,.
\end{equation}
This matrix controls how the quadric depends on the active integration variables $x_i$ and $x_j$: $(XQX) = q_{ii}x_i^2+2q_{ij}x_ix_j+q_{jj}x_j^2 + \text{rest}$.
Consequently, the first symbol entry $\overline{ij}_{\text{(first entry)}}$ associated with this choice is controled by the matrix $Q_{(ij)}$:
\begin{equation}\label{eq:firstentry}
    \overline{ij}_{\text{first entry}}=\left\{
    \begin{matrix}
        &r\left(Q_{(ij)}^{-1}\right),\quad &q_{ii}\neq0,\,q_{jj}\neq0,\\
        &\left(\frac{q_{ij}^2}{q_{jj}}\right)^{-\mathrm{sign}(q_{ij})},\quad &q_{ii}=0,\, q_{jj}\neq0,\\
        &\left(\frac{q_{ij}^2}{q_{ii}}\right)^{-\mathrm{sign}(q_{ij})},\quad &q_{ii}\neq0,\, q_{jj}=0,\\
        &q_{ij}^{-2\mathrm{sign}(q_{ij})},\quad &q_{ii}=0,\, q_{jj}=0,\\
    \end{matrix}
    \right.
\end{equation}
where $r$ is a function of a $2\times2$ matrix 
\begin{equation}
    r(M_{2\times2})=\frac{M_{12}-\sqrt{M_{12}^2-M_{11}M_{22}}}{M_{12}+\sqrt{M_{12}^2-M_{11}M_{22}}}. 
\end{equation}
This implies that $I$ has a branch cut starting at $\overline{ij}_{\mathrm{first\ entry}}=0$ and ending at ${1/\overline{ij}_{\mathrm{first\ entry}}}=0$.

\paragraph{Discontinuities.} 
Next, we compute the discontinuity associated to the $\overline{ij}_{\mathrm{first\ entry}}=0$ and ${1/\overline{ij}_{\mathrm{first\ entry}}}=0$ branch cut via a shpherical contour.

To achieve this, we make a change of variables $\{x_i,x_j\}\mapsto\{w_i,w_j\}$ so that the quadric takes the form
\begin{equation} \label{eq:lightConeQuadric}
    w_iw_j+X_{\widehat{\{i,j\}}}Q^{(ij)}X_{\widehat{\{i,j\}}}.
\end{equation}
Note that $Q^{(ij)}$ is a new quadric ($Q^{(ij)} \neq Q_{(ij)}$) that becomes the divisor of the new integral after integrating over $w_i$ and $w_j$; this new integral is, of course, the $\overline{ij}$-discontinuity of $I$ (or one of its sequential discontinuities). 
The wide hat $\widehat{\{i,j\}}$ instructs us to forget about the variables $x_i$ and $x_j$
\begin{equation}
    X_{\widehat{\{i,j\}}}=[x_1 : \cdots : x_{i-1} : x_{i+1} : \cdots : x_{j-1} : x_{j+1} : \cdots : x_{n}].
\end{equation} 
We also use $Q_{\{i,j\},\widehat{\{i,j\}}}$ ($Q_{\widehat{\{i,j\}},\{i,j\}}$) to denote the $2\times(n-2)$ ($(n-2)\times2$) matrix made up of the $i$'th and $j$'th rows (columns) of $Q$ with the $i$'th and $j$'th columns (rows) deleted. 
Using the above notation, the change of variables $\{x_i,x_j\}\mapsto\{w_i,w_j\}$ is 
\begin{equation}
    \left(\begin{matrix}
        x_i\\
        x_j
    \end{matrix}\right)=R\left(\begin{matrix}
        w_i\\
        w_j
    \end{matrix}\right)
    -  Q_{(ij)}^{-1} 
    Q_{ \{i,j\}, \widehat{\{i,j\}} }
    X_{\widehat{\{i,j\}}}
    \,,
\end{equation}
where $R$ is a $2\times2$ matrix constrained to satisfy 
\begin{align}
    R^T 
     Q_{(ij)} 
    R
    = \left(\begin{matrix}
        0&1\over2\\
        1\over2& 0
    \end{matrix}\right).
    \label{eq:RmatEq}
\end{align}
In fact, there are two solutions to $R$, and one can choose either one.
The new quadric $Q^{(ij)}$ is
\begin{equation}\label{eq:projQij}
    Q^{(ij)}=Q_{\widehat{\{i,j\}},\widehat{\{i,j\}}}-Q_{\widehat{\{i,j\}},\{i,j\}}Q^{-1}_{\{i,j\},\{i,j\}}Q_{\{i,j\},\widehat{\{i,j\}}}
    \,,
\end{equation}
where $Q_{\widehat{\{i,j\}},\widehat{\{i,j\}}}$ is the $(n-2)\times(n-2)$ matrix obtained from $Q$ after deleting the $i$'th and $j$'th rows as well as the $i$'th and $j$'th columns.

After the above change of variables, the integrand in  \eqref{eq:Idef} becomes
\begin{align}
    \frac{\langle X\mathrm{d}X^{n-1}\rangle\,T[X^k]}{(XQX)^{\frac{n+k}{2}}} = 
    \sum_{p_i,p_j} t_{p_ip_j}
    \langle X_{\widehat{\{i,j\}}} \mathrm{d} X_{\widehat{\{i,j\}}}^{n-3} \rangle 
    \frac{
        \mathrm{d}w_i\mathrm{d}w_j\; w_i^{p_i}w_j^{p_j}
    }{
        (w_iw_j+X_{\widehat{\{i,j\}}}Q^{(ij)}X_{\widehat{\{i,j\}}})^{n+k\over2}
    }
    \,.
\end{align}
where $t_{p_i p_j}$ is a degree-$(k-p_i-p_j)$ polynomial of the remaining integration variables $ X_{\widehat{\{i,j\}}}$. 
We have also absorbed the coefficients of the tensor structure $T$ and the Jacobian into the $t_{p_i p_j}$. 
We can now perform the spherical contour $S^2$ integral by setting $w_i=r\,e^{i\phi}$ and $w_j=r\,e^{-i\phi}$,
and, integrating over the region where $r\in[0,+\infty]$ and $\phi\in[0,2\pi]$. 
Then, 
\begin{align}\begin{aligned}
    \mathrm{Disc}_{\overline{ij}}[I] := &
    \sum_{p_i,p_j}
    \int_{\Delta^{(ij)}} 
    \langle 
        X_{\widehat{\{i,j\}}} 
        \mathrm{d}
        X_{\widehat{\{i,j\}}}^{n-3} 
    \rangle 
    C_{p_ip_j} 
    \\[-2em] & \hspace{4em} \times 
        \int_{0}^{\infty}
        \frac{
            \mathrm{d}r ({-}2ir) \; 
            r^{p_i{+}p_j}
        }{
            (
                r^2
                {+} X_{\widehat{\{i,j\}}} 
                    Q^{(ij)} 
                    X_{\widehat{\{i,j\}}}
            )^{n{+}k\over2}
        }
        \;
        \overset{2\pi\delta_{p_ip_j}}{\overbrace{
            \int_{0}^{2\pi} \mathrm{d}\phi \;
            e^{i \phi(p_i{-}p_j)}
        }}
    \,.
\end{aligned}\end{align}
where $\Delta^{(ij)}$ is a standard $(n-3)$-simplex for $X_{\widehat{\{i,j\}}} \in \mathbb{CP}^{n-3}$.
Note that the $\phi$-integral forces $p_i=p_j$.
Performing the radial integration yields an integral of the same form as \eqref{eq:Idef}
\begin{equation}
    \mathrm{Disc}_{\overline{ij}}[I] =
    \int_{\Delta^{(ij)}} 
    \frac{
        \langle 
            X_{\widehat{\{i,j\}}} 
            \mathrm{d} X_{\widehat{\{i,j\}}} ^{n'-1}
        \rangle 
        T^\prime[X_{\widehat{\{i,j\}}}^k]
        }{
            (
                X_{\widehat{\{i,j\}}} 
                Q^{(ij)} 
                X_{\widehat{\{i,j\}}}
            )^{n'+k\over2}
        }
    \,,
\end{equation}
where $n'=n-2$ and $T^\prime[X^k] = (-i) \sum_{p=0}^k \frac{\Gamma(p+1)\Gamma({k+n\over2}-p-1)}{\Gamma({k+n\over2}) } t_{pp} (X_{\widehat{\{i,j\}}} Q^{(ij)} X_{\widehat{\{i,j\}}})^p$.

\paragraph{Coefficients.} 
After performing $\lfloor \frac{n-1}{2}\rfloor$ spherical contours, one is left with an integrand that is either a one-form (even $n$) or a zero-form integral (odd $n$). 

If $n$ is even, performing $\lfloor \frac{n-1}{2}\rfloor$ spherical contours turns the original $(n-1)$-form into a 1-form proportional to the volume element $\la \tilde{X} \d\tilde{X}^1\ra = \tilde{X}_1\; \d \tilde{X}_2 - \tilde{X}_2\; \d \tilde{X}_1$ where $\tilde{X}$ is the remaining integration variables in $\mathbb{CP}^1$. 
To extract the coefficient, make the variable transformation that puts the quadric into the form of equation \eqref{eq:lightConeQuadric} then change to radial coordinates $\tilde{w}= r e^{i\theta}$ and $\overline{\tilde{w}} = r e^{-i\theta}$
\begin{align} \label{eq:even}
    \la \tilde{X} \d \tilde{X}^1 \ra 
    \propto \tilde{w}\; \d \overline{\tilde{w}} - \overline{\tilde{w}}\; \d \tilde{w}
    \to -2ir\; \d\theta
    \,.
\end{align}
The lack of a $\d r$ component indicates that only angular part of the spherical contour is needed to extract the coefficient. 
Due to projective invariance, all $r$-dependence cancels (i.e., one can safely set $r=1$). 
The result of these spherical contours is the coefficient after multiplying by $\left(\frac{1}{2 (2\pi i)}\right)^{\lfloor\frac{n-1}{2}\rfloor}$ (c.f., \eqref{eq:coeffs}). 

When $n$ is odd, there is no integration left:
taking $\lfloor \frac{n-1}{2}\rfloor$ spherical contours turns the original $(n-1)$-form into a 0-form. 
The resulting 0-form is a function of the final $x_k$. 
However, because of projective invariance, all dependence on $x_k$ cancels (i.e., one can safely set $x_k=1$). 
The coefficient is the result of these spherical contours multiplied by $\left(\frac{1}{2 (2\pi i)}\right)^{\lfloor\frac{n-1}{2}\rfloor}$ (c.f., \eqref{eq:coeffs}). 

The factors of $\frac{1}{2}$ are needed because of how $\overline{ij}$ is defined in equation \eqref{eq:firstentry}.
To see this, it is simplest to look at the case where $q_{ii}=q_{jj}=0$ where $\overline{ij} = q_{ij}^{-2\mathrm{sign}(q_{ij})}$. 
The spherical contour $S_{ij}^2$ extracts the discontinuity across the logarithmic branch cut stretching between $1/q_{ij}=0$ and $q_{ij}=0$. 
Because of how the argument of the logarithm is represented (i.e., $\overline{ij}$) we must compensate the spherical contour by a factor of $\frac{1}{2}$ since $\log q_{ij}^{-2\mathrm{sign}(q_{ij})} = 2 \log q_{ij}^{-\mathrm{sign}(q_{ij})}$. 
Moreover, each $\d\theta$ integration produces a factor of $2\pi i$ that should not be associated with the coefficient $C_{\bs{\alpha}}$ and one also has to divide by $2\pi i$ for each spherical contour.

Both even/odd-$n$ cases appear in the $\mathcal{N}=4$ SYM and QCD bootstrap. 
In the next section, we provide fully worked examples of the spherical contour algorithms in the context of $\mathcal{N}=4$ SYM EEEC.

\subsubsection{Examples from $\mathcal{N}=4$ EEEC}
\label{sec:examples}

In this section, we provide several examples for obtaining the leading transcendentality part of $\mathcal{N}=4$ SYM EEEC using the spherical contour method. 
In $\mathcal{N}=4$ SYM case, the EEEC can be expressed in terms of two integrals
\begin{align}
    F_1&=3\int_{\Delta_4} {\la Z\rm{d}Z^3\ra} \times \frac{z_2 z_3 z_4^2}{\left[ z_1 z_2+(z\bar z) z_2z_3 +(1-z)(1-\bar z) z_1 z_3 +(z_1+z_2+z_3)z_4 \right]^4}\,,\\
    F_2&=12\int_{\Delta_5} {\la Z \rm{d}Z^4\ra} \frac{z_1 z_2 z_3 z_5^2}{\left[ z_1 z_2+(z\bar z) z_2z_3 +(1-z)(1-\bar z) z_1 z_3 +(z_2+z_3)z_4 +(z_1+z_2+z_3)z_5\right]^5}\,.\nn
\end{align}
where $F_1$ contains four Feynman-like parameters and $F_2$ has five Feynman-like parameters. 
We treat $F_1$ in detail first and then demonstrate the spherical contour algorithm in the context of $F_2$. 

Note that in the examples we divide $4\pi i$ for each spherical contour instead of waiting till the end of the calculation as in \eqref{eq:coeffs}.

\paragraph{Example 1: $F_1$.}
We begin with the first integral $F_1$, where the quadric is
\begin{equation}
    Q_1=\frac{1}2{}\left(\begin{matrix}
        0 &1 &(1-z)(1-\bar{z})&1\\
        1 &0 & z \bar{z}& 1\\
        (1-z)(1-\bar{z}) &z\bar{z}& 0& 1\\
        1&1&1&0
    \end{matrix}\right).
\end{equation}
There are only two possible first entries $\overline{13}$ and $\overline{23}$ that are not constant and contribute to the symbol. 
We demonstrate the contribution to the symbol form the spherical contours $S^2_{14} \circ S^2_{23}$ which is related to sequential discontinuities $\mathrm{Disc}_{b=0}\circ\mathrm{Disc}_{a=0}$ where $b\in\{ (1-z),(1-\overline{z})\}$ and $a\in\{z,\bar{z}\}$. 

The $\overline{13}$ first entry, according to \eqref{eq:firstentry}, is $(z \bar{z})^{-2\mathrm{Sign}(z\bar{z})}$. 
Then we perform a change of variable by solving the equation
\begin{equation}
    R^\mathrm{T}(Q_1)_{\{2,3\}}R=\left(\begin{matrix}
        0&1\over2\\
        1\over2&0
    \end{matrix}\right)\Longrightarrow R_1=\left(\begin{matrix}
        c&0\\
        0& 1\over c\  z \bar{z}
    \end{matrix}\right),\,R_2=\left(\begin{matrix}
        0&c\\
        1\over c\,z \bar{z}& 0
    \end{matrix}\right)
\end{equation}
where $c$ is any constant, which means there is one degree of freedom. One can choose any of the solutions and set $c$ to a suitable number. Here we choose $R_1$ and set $c=1$. The change of variables now become
\begin{equation}
    \left(\begin{matrix}
        z_2\\z_3
    \end{matrix}\right)=R\left(\begin{matrix}
        w_2\\w_3
    \end{matrix}\right)-\left(Q_1\right)_{\{2,3\},\{2,3\}}^{-1}(Q_1)_{\{2,3\},\{1,4\}}\left(\begin{matrix}
        z_1\\z_4
    \end{matrix}\right)=\left(\begin{matrix}
        w_2-\frac{(1-z)(1-\bar{z})}{z\bar{z}}z_1-\frac{1}{z\bar{z}}z_4\\{(w_3-z_1-z_4)}/{z\bar{z}}
    \end{matrix}\right),
\end{equation}
and the Jacobian is $1/z\bar{z}$. Now one can take the spherical contour to compute the residue by taking $w_2=r e^{-i\theta}, w_3=r e^{i\theta}$, and the discontinuity corresponding to $z\bar{z}=0$ is
\begin{equation}
    \mathrm{Disc}_{\overline{23}}F_1={{\frac{3}{4\pi i}}}\int\langle Z\mathrm{d}Z\rangle\int_0^\infty\mathrm{d}r \,(-2ir)\int_0^{2\pi}\mathrm{d}\phi\frac{1}{z\bar{z}}\frac{N[r,\phi;z_1,z_4]}{\left(r^2+Z_{\{1,4\}}Q_1^{(2,3)}Z_{\{1,4\}}\right)^4},
\end{equation}
where the new quadric matrix is
\begin{equation}\label{eq:ex1quad23}
    Q_1^{(2,3)}=\frac{1}{2z\bar{z}}\left(\begin{matrix}
        -2(1-z)(1-\bar{z})&z+\bar{z}-2\\
        z+\bar{z}-2&-2
    \end{matrix}\right),
\end{equation}
and the numerator is now
\begin{align}
    N[r,\phi;z_1,z_4]=\frac{1}{(z\bar{z})^3}&\left((z\bar{z})^2r^2z_4^2-r(e^{-i\theta}(z\bar{z})^2z_4^2(z_1+z_4)-e^{i\theta}(z\bar{z})^2z_4^2((1-z)(1-\bar{z})w_1+w_4))\right.\nn\\
    &\,\,\,+\left.(z\bar{z})z_4^2(z_1+z_4)((1-z)(1-\bar{z})w_1+w_4))\right).
\end{align}
The integration over $\theta$ is trivial since only terms without an exponential survive.
After integration over $r$ and $\theta$, the discontinuity is
\begin{equation}\label{eq:disc23F1}
    \mathrm{Disc}_{\overline{23}}F_1=-\frac{1}{4(z\bar{z})^3}\int\frac{z_4^2~\langle Z\mathrm{d}Z\rangle}{(Z_{\{1,4\}}Q_1^{(2,3)}Z_{\{1,4\}})^3}\left((1-z)(1-\bar{z})z_1^2+z_4^2+(2-z-\bar{z}+2z\bar{z})z_1z_4\right).
\end{equation}
The second entry can be calculated in the same way from \eqref{eq:ex1quad23}, and this symbol term becomes 
\begin{equation}\label{eq:scsqrt}
    C_{23,14} \,\left((z\bar{z})^{-2\mathrm{Sign(z\bar{z})}}\otimes\frac{1-z}{1-\bar{z}}\right)
    \subset \mathcal{S}[F_1]
    ,
\end{equation}
where $
    \overline{14}_{Q^{(2,3)}_1} 
    = \frac{
        -2+z+\bar{z}+\sqrt{(z-\bar{z})^2}
    }{
        2-z+\bar{z}+\sqrt{(z-\bar{z})^2}
    } 
    = \frac{1-z}{1-\bar{z}}
$.\footnote{%
    We put the matrix $Q_1^{(2,3)}$ in the subscript of $\overline{14}$ to make it clear that the first entry is computed with respect to the matrix $Q_1^{(2,3)}$. 
} 
Note that here we choose the real region $z>\bar{z}$ to fix the sign of the square root and analytically continue to the general region. If there are more square roots, one should choose a suitable region to cancel them in a unified region. To extract the coefficient $C_{23,14}$, we act with the spherical contour $S^2_{14}$ on \eqref{eq:disc23F1}. 
To do this we need a variable change \eqref{eq:RmatEq} such that 
\begin{align}
    (2z \bar{z})R^{\mathrm{T}}Q_1^{(2,3)}R&=\left(\begin{matrix}
        0&1\over2\\
        1\over2&0
    \end{matrix}\right)
    \,.
\end{align}
One convenient choice for $R$ is 
\begin{align}
    R=\left(\begin{matrix}
        1&-\frac{1}{(z-\bar{z})^2}\\
        -(1-z)& \frac{1-\bar{z}}{(z-\bar{z})^2}
    \end{matrix}\right)
    ,
\end{align}
and the transformation becomes
\begin{equation}
    \left(\begin{matrix}
        z_1\\z_4
    \end{matrix}\right)=R\left(\begin{matrix}
        w_1\\w_4
    \end{matrix}\right)=\left(\begin{matrix}
        w_1-\frac{1}{(z-\bar{z})^2}w_4\\-(1-z)w_1+\frac{1-\bar{z}}{(z-\bar{z})^2}w_4
    \end{matrix}\right).
\end{equation}
with Jacobian $\frac{1}{\sqrt{(z-\bar{z})^2}}=\frac{1}{z-\bar{z}}$. 
We perform the spherical contour by setting $w_1=re^{-i\theta},w_4=re^{i\theta}$, and after integration over $\theta$ variable, we get
\begin{equation}
    C_{23,14} = \mathrm{Disc}_{\overline{14}_{Q_1^{(2,3)}}} \mathrm{Disc}_{\overline{23}}F_1 
    = -\frac{(1-z)(1-\bar{z})}{2(z-\bar{z})^5}(-z^2-\bar{z}^2-4z \bar{z}+3z^2\bar{z}+3z\bar{z}^2).  
\end{equation}
This is the even case as illustrated in \eqref{eq:even}.
Thus, the $S^2_{14}\circ S^2_{23}$ spherical contour fixes the following terms in the symbol
\begin{equation}
    C_{23,14}
    \,\left(\frac{1}{z^2\bar{z}^2}\otimes\frac{1-z}{1-\bar{z}}\right) 
    \subset \mathcal{S}[F_1]
    ,
\end{equation}
which matches the known expression for $F_1$. 
Note that if we choose another real region $z<\bar{z}$ in \eqref{eq:scsqrt}, the coefficient will give an extra minus sign. However, the second symbol entry will become its inverse and contribute to another minus sign, which will give the same result. 
A similar story holds for the iterated spherical contour $S^2_{24}\circ S^2_{13}$. 
The iterated spherical contours $S^2_{14}\circ S^2_{23}$ and $S^2_{24}\circ S^2_{13}$ completely fix the leading transcendental weight part of $\mathcal{S}(F_1)$.

\paragraph{Example 2: $F_2$.} Consider the example $F_2$ which has the associated quadric 
\begin{equation}
    Q_2=\frac{1}{2}\left(\begin{matrix}
        0&1&(1-z)(1-\bar{z})&0&1\\
        1&0&z\bar{z}&1&1\\
        (1-z)(1-\bar{z})&z\bar{z}&0&1&1\\
        0&1&1&0&0\\
        0&1&1&0&0\\
    \end{matrix}\right).
\end{equation}
Like $F_1$, there are only two non-trivial first entries $\overline{13}$ and $\overline{23}$. 
However, after performing the spherical contour of the first entry, the new quadric is dimension $3$, and thus contributes to $3$ (or fewer) second entries. 
To illustrate the spherical contour algorithm in the context of $F_2$, we choose to act with $S^2_{13}$ and list the important steps instead of a complete discussion. 

The first entry $\overline{13}$ is $((1-z)(1-\bar{z}))^{-2}$, and its related spherical contour change of variable is
\begin{equation}
    \left(\begin{matrix}
        z_1\\z_3
    \end{matrix}\right)=\left(\begin{matrix}
        w_1-\frac{z\bar{z}z_2+z_4+z_5}{(1-z)(1-\bar{z})}\\\frac{w_3-z_2-z_5}{(1-z)(1-\bar{z})}
    \end{matrix}\right),
\end{equation}
with Jacobian $J=\frac{1}{(1-z)(1-\bar{z})}$. 
After performing the spherical contour integration by setting $w_1=re^{i\theta},w_3=re^{-i\theta}$, the discontinuity becomes
\begin{align}
    \mathrm{Disc}_{\overline{13}}F_2&=\int\langle Z\mathrm{d}Z^2\rangle\\
    \times&\frac{(1-z)(1-\bar{z})z_2z_5^2(((-3+\bar{z})z_2-2z_5)(z_4+z_5)+zz_2(z_4+z_5-\bar{z}(2z_2+z_4+3z_5)))}{((\bar{z}z_2+z_5)(z_4+z_5)+z z_2(\bar{z}(z_2-z_4)+z_4+z_5))^4}.
    \nn
\end{align}
Then, the new quadric becomes
\begin{equation}
    Q_2^{(1,3)}=\frac{1}{2}\left(\begin{matrix}
        2z\bar{z}&z+\bar{z}-z\bar{z}&z+\bar{z}\\
        z+\bar{z}-z\bar{z}&0&1\\
        z+\bar{z}&1&2
    \end{matrix}\right).
\end{equation}
There are two second entries $\overline{24}_{Q_2^{(1,3)}}$ and $\overline{24}_{Q_2^{(2,5)}}$ from the above quadric. 

Next, we act with the spherical contour $S^2_{25}$ with  corresponding second entry is $\overline{24}_{Q_2^{(2,5)}} = \bar{z}/z$. 
To perform the $S^2_{25}$ spherical contour, we make the variable transformation
\begin{equation}
    \left(\begin{matrix}
        z_2\\z_5
    \end{matrix}\right)=\left(\begin{matrix}
        w_2-\frac{w_5}{2(z-\bar{z})^2}-\frac{2z\bar{z}-z-\bar{z}}{(z-\bar{z})^2}z_4\\
        -zw_z+\frac{\bar{z}w_5}{2(z-\bar{z})^2}+\frac{z\bar{z}(z+\bar{z})-z^2-\bar{z}^2}{z-\bar{z}}
    \end{matrix}\right),
\end{equation}
with Jacobian $J=\frac{1}{2(z-\bar{z})}$ and set $w_2=re^{i\theta}, w_5=re^{-i\theta}$ (we assume $z>\bar{z}$ to fix the sign).
With this variable change $\la Z \d Z^2\ra = z_4 (-2ir) \d r\wedge \d\theta$. 
Performing the $r$- and $\theta$-integration, yields the coefficient $C_{13,25}$
\begin{align}
    C_{13,25} 
    = \mathrm{Disc}_{\overline{25}_{Q_2^{(1,3)}}}
    \mathrm{Disc}_{\overline{13}}
    & F_2
    =
    \frac{1}{4z^2\bar{z}^2(z-\bar{z})^5}\nn\\
    \times(&z^6(1-\bar{z})+\bar{z}^6+5z^2\bar{z}^4(1+\bar{z})-z\bar{z}^5(4+\bar{z})+2z^3\bar{z}^3(4-11\bar{z}+2\bar{z}^2)\nn\\
    &+z^4\bar{z}^2(5-22\bar{z}+28\bar{z}^2-6\bar{z}^3)+z^5\bar{z}(-4+5\bar{z}+4\bar{z}^2-6\bar{z}^3))
\end{align}
Therefore, the $S^2_{25}\circ S^2_{13}$ fixes the following term in the symbol of $F_2$
\begin{equation}
    C_{13,25}
    \left(
        \frac{1}{(1-z)^2(1-\bar{z})^2}\otimes\frac{\bar{z}}{z}
    \right)
    \subset \mathcal{S}[F_2].
\end{equation}
Similar operations can be performed on other orders, and one can completely calculate its leading transcendental weight symbol by the spherical contour.

\section{Bootstrapping three-point energy correlators}
\label{sec:bootstrap3}
With all the tools at hand, we now study how to determine the analytic expressions of collinear three-point energy correlators at LO using spherical contour, physical constraints, and squeezed limit data. In the following, we will calculate the results for $\mathcal{N}=4$ SYM as well as quark and gluon jets. In particular, we decompose both quark and gluon jets into the partonic channels listed in Eq.~\eqref{eq:qcdchannel}. 

In this section, we focus on the bootstrap framework for $a=b=c=1$ correlators; however, the method can be applied to any energy weights as long as $a,b,c\geq 1$. 
In the next section, we introduce an alternative method to replace the constraints from squeezed limit data and explicitly work out several higher energy weights.

\subsection{Workflow}
\begin{figure}[!htbp]
    \centering
    \includegraphics[scale=0.4]{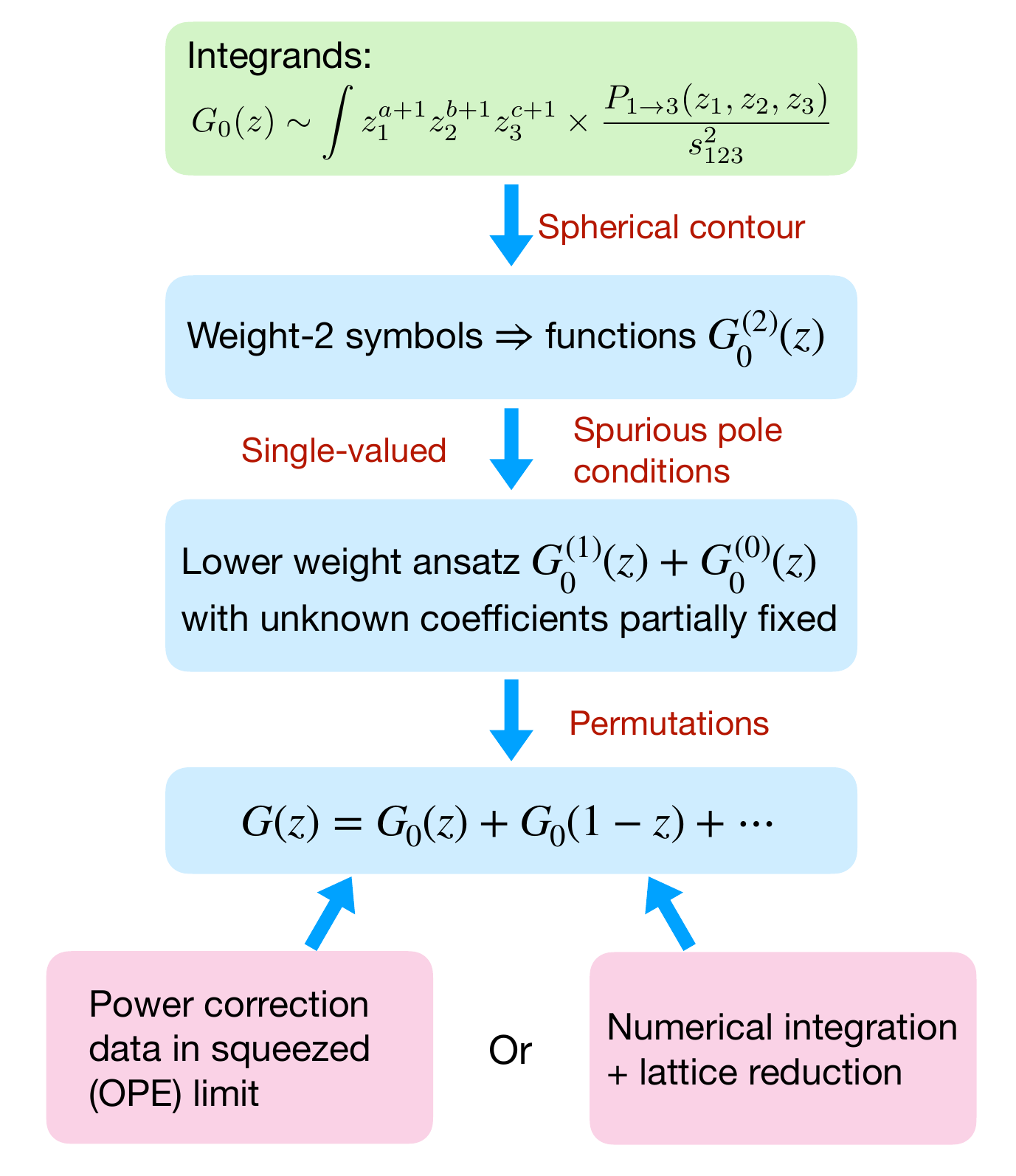}
    \caption{Workflow for bootstrap three-point energy correlators. This illustrates how to obtain the full analytic expression using spherical contour, single-valued condition, spurious pole conditions, as well as the physical data, from either the squeezed (OPE) limit or numerical integration.}
    \label{fig:workflow}
\end{figure}

Firstly, we explain the workflow of the bootstrap program for three-point correlators. A step-by-step procedure is summarized in Fig.~\ref{fig:workflow}. 

As described in Sec.~\ref{sec:factorization}, the collinear limit allows us to start with the integrand $G_0(z)$ defined in Eq.~\eqref{eq:G0} for splitting channels. Using the spherical contour method discussed in Sec.~\ref{sec:spherical_contour}, we can then determine the weight-2 symbol expression from the singularity structure of the integrand. Such symbols already satisfy the integrable conditions and more impressively, the spherical contour extracts the rational coefficients in the weight two symbols. Using the function \texttt{FiberSymbol} in the package \textsc{PolyLogTools}~\cite{Duhr:2019tlz}, we can then integrate these symbols into polylogarithmic functions.

The complete expression of the transcendental weight-2 part provides us with some insights for lower weights. As we will see below, the weight-2 part is expressed in terms of single-valued polylogarithms, which implies that the lower-weight part must also be single-valued. Secondly, the rational coefficients in weight-2 contain some spurious poles. The fact that these poles have to cancel constrains the coefficients in the lower weights. With this information, we can write down the ansatz for both the weight-1 and weight-0 terms as well as partially determine the unknown parameters from the cancellation of spurious poles. There are additional cancellations after summing over all six permutations of $G_0(z)$, further reducing the number of unknown parameters. 

Finally, we provide two different ways to fix the rest of the parameters. One way is to make use of the expansion in the squeezed limit $z\to 0$, $z\to 1$ or $z\to \infty$. Take $z\to 0$ as an example, the leading power $1/(z\bar z)$ is predicted by the factorization formula introduced in Ref.~\cite{Chen:2019bpb}, while higher power correction terms $(z \bar z), (z\bar z)^2, \cdots$ come from power expanding the integrand and performing the integration. Ref.~\cite{Chen:2021gdk} explains the computation in detail and makes connections to the light-ray OPE. We refer to it as the {\it OPE approach} and we explain it in this section. Alternatively, we numerically integrate the three-point correlators to high precision and perform an analytic regression to determine the unknown parameters. As studied in Ref.~\cite{Barrera:2025uin}, this can be done using the lattice reduction method. We will call this the {\it lattice reduction approach} and discuss it in the next section. 

\subsection{Leading transcendentality}

The spherical contour method extracts both the symbols and their corresponding coefficients of the integrals with quadratic denominators. 
To begin with, we need to select all the quadratic integrands from the EEEC integrand and turn them into Feynman parameter integrals. 
For example, in the $\mathcal{N}=4$ SYM splitting in Eq.~\eqref{eq:N4_split}, divided by the $s_{123}^2$ in Eq.~\eqref{eq:G0}, the first two terms only contain linear denominators and will not give rise to transcendental weight-2 functions. Then we are left with
\begin{equation}
    \int \d z_1 \d z_2 \d z_3\, \delta(1-z_1-z_2-z_3) z_1^{a+1} z_2^{b+1} z_3^{c+1}\times \frac{1}{s_{123}s_{12}z_3}\left(\frac{1}{z_1}+\frac{1}{z_2+z_3}\right)\,.
\end{equation}
For $a=b=c=1$, we arrive at two integrals.
\begin{align}
    F_1&=\int_0^\infty \d z_1 \d z_2 \d z_3\, \delta(1-z_1-z_2-z_3)\times\frac{z_2 z_3}{z_1z_2+(z\bar z) z_2 z_3+(1-z)(1-\bar z)z_1 z_3}\,,
    \\
    F_2&=\int_0^\infty \d z_1 \d z_2 \d z_3\, \delta(1-z_1-z_2-z_3)\times\frac{z_1 z_2 z_3}{z_1z_2+(z\bar z) z_2 z_3+(1-z)(1-\bar z)z_1 z_3}\frac{1}{z_2+z_3}\,.
    \nn
\end{align}
The goal of Feynman parameterization is to make the integrands invariant under uniform scaling of the Feynman parameters, as required by the spherical contour. In particular, the GL(1) transformation will allow us to pick the most convenient contour. For $F_1$, we introduce one more Feynman parameter $z_4$ and rewrite it as
\begin{align}
    F_1&=3 \int_0^\infty \frac{\d z_1 \d z_2 \d z_3 \d z_4}{\text{GL}(1)} \times \frac{z_2 z_3 z_4^2}{\left[ z_1 z_2+(z\bar z) z_2z_3 +(1-z)(1-\bar z) z_1 z_3 +(z_1+z_2+z_3)z_4 \right]^4}\,\nn,\\
    &=3\int_{\Delta_4} {\la Z\rm{d}Z^3\ra} \times \frac{z_2 z_3 z_4^2}{\left[ z_1 z_2+(z\bar z) z_2z_3 +(1-z)(1-\bar z) z_1 z_3 +(z_1+z_2+z_3)z_4 \right]^4}\,.
\end{align}
For $F_2$, since we need to combine the two denominators into one single denominator, we will need two more Feynman parameters.
\begin{align}
    F_2&=\frac{\Gamma(5)}{\Gamma(2)\Gamma(3)} \int_0^\infty \frac{\d z_1 \d z_2 \d z_3 \d z_4 \d z_5}{\text{GL}(1)}\\ 
    &\qquad\qquad\qquad~ \times \frac{z_1 z_2 z_3 z_5^2}{\left[ z_1 z_2+(z\bar z) z_2z_3 +(1-z)(1-\bar z) z_1 z_3 +(z_2+z_3)z_4 +(z_1+z_2+z_3)z_5\right]^5}\,\nn,\\
    &=12\int_{\Delta_5} {\la Z \rm{d}Z^4\ra} \frac{z_1 z_2 z_3 z_5^2}{\left[ z_1 z_2+(z\bar z) z_2z_3 +(1-z)(1-\bar z) z_1 z_3 +(z_2+z_3)z_4 +(z_1+z_2+z_3)z_5\right]^5}\,.\nn
\end{align}
The detailed calculations of $F_{1,2}$ using spherical contours can be found in Sec.~\ref{sec:examples}. 

For QCD splittings, we apply the same procedure: select the quadratic integrand and perform the Feynman parameterization. For a quark jet, we get 17 terms for $P_{qq^\prime \bar q^\prime}$, 4 terms for $P_{qq\bar q}$, 7 terms for $P_{qgg}^{(C_F)}$ and 28 terms for $P_{qgg}^{(C_A)}$; for a gluon jet, we have 7 terms for $P_{gq\bar q}^{(C_F)}$, 23 terms for $P_{gq\bar q}^{(C_A)}$ and 22 terms for $P_{ggg}$. With the single-denominator form at hand, we can use the spherical contour approach to compute the symbol expressions and the residue for each integrand. The symbol expressions are then integrated to weight-2 polylogarithms with \texttt{FiberSymbol} in \textsc{PolyLogTools}~\cite{Duhr:2019tlz}, since all entries are linear in $z$ and $\bar z$. Note that we have not imposed any boundary conditions, so the $\pi^2$ term obtained from \texttt{FiberSymbol} could be artificial. We will fix them in the next steps.
For $a=b=c=1$, we also verify the result with the analytic expressions in Ref.~\cite{Chen:2019bpb}. 

For illustration, here we quote the result for $P_{q q^\prime \bar q^\prime}$ with color factor $C_F T_F n_f$ in the quark jet channel:
\begin{align}
    G_0^{(2)}(z)&\supset K= \frac{1}{48} \pi ^2 \left(-z^3-z^2 (\bar z-2)-z \left(\bar z^2-2 \bar z+2\right)-\bar z^3+2 \bar z^2-2 \bar z+1\right)\nn\\
    &+\frac{1}{8} \left(z^3+z^2 (\bar z-2)+z \left(\bar z^2-2
   \bar z+2\right)+\bar z^3-2 \bar z^2+2
   \bar z-1\right) \nn\\
   &\times\Bigg[\text{Li}_2\left(1-\frac{z
   \bar z}{(1-z) (1-\bar z)}\right)+\frac{1}{2} \log
   \left(\frac{1}{(1-z) (1-\bar z)}\right) \log \left(\frac{z
   \bar z}{(1-z) (1-\bar z)}\right)\Bigg]\nn\\
   &+\frac{i D_2^{-}(z)}{4(z-\bar z)^{11}}\Bigg[z^{14}-2 z^{13} (5 \bar z+1)+z^{12} \left(45 \bar z^2+20
   \bar z+2\right)-z^{11} \left(120 \bar z^3+90
   \bar z^2+20 \bar z+1\right)\nn\\
   &+11 z^{10} \bar z \left(19
   \bar z^3+22 \bar z^2+8 \bar z+1\right)-11 z^9
   \bar z^2 \left(22 \bar z^3+40 \bar z^2+20
   \bar z+5\right)\nn\\
   &+z^8 \bar z \left(-675 \bar z^5+3114
   \bar z^4-2495 \bar z^3+1615 \bar z^2-330
   \bar z+25\right)\nn\\
   &+z^7 \bar z \left(-1440 \bar z^6+7740
   \bar z^5-16838 \bar z^4+14605 \bar z^3-6500
   \bar z^2+1255 \bar z-76\right)\nn\\
   &+z^6 \bar z \left(-675
   \bar z^7+7740 \bar z^6-26050 \bar z^5+41281
   \bar z^4-31230 \bar z^3+11695 \bar z^2-1889
   \bar z+85\right)\nn\\
   &+z^5 \bar z \left(-242 \bar z^8+3114
   \bar z^7-16838 \bar z^6+41281 \bar z^5-52400
   \bar z^4+33897 \bar z^3-10674 \bar z^2+1348
   \bar z-36\right)\nn\\
   &+z^4 \bar z^2 \left(209 \bar z^8-440
   \bar z^7-2495 \bar z^6+14605 \bar z^5-31230
   \bar z^4+33897 \bar z^3-18570 \bar z^2+4615
   \bar z-360\right)\nn\\
   &+z^3 \bar z^3 \left(-120
   \bar z^8+242 \bar z^7-220 \bar z^6+1615
   \bar z^5-6500 \bar z^4+11695 \bar z^3-10674
   \bar z^2+4615 \bar z-720\right)\nn\\
   &+z^2 \bar z^4 \left(45
   \bar z^8-90 \bar z^7+88 \bar z^6-55 \bar z^5-330
   \bar z^4+1255 \bar z^3-1889 \bar z^2+1348
   \bar z-360\right)\nn\\
   &+z \bar z^5 \left(-10 \bar z^8+20
   \bar z^7-20 \bar z^6+11 \bar z^5+25 \bar z^3-76
   \bar z^2+85 \bar z-36\right)+\bar z^{11}
   \left(\bar z^3-2 \bar z^2+2 \bar z-1\right)\Bigg]\,,
   \label{eq:LT_nf}
\end{align}
with Bloch-Wigner function
\begin{equation}
    2i D_2^-(z) = \text{Li}_2(z) - \text{Li}_2(\bar{z}) + \frac{1}{2} \left( \log(1 - z) - \log(1 - \bar{z}) \right) \log(z \bar{z}) \, .
\end{equation}
Defining a parity-even function:
\begin{equation}
    D_{2}^{+}(z)=\text{Li}_2(1-z\bar z) +\frac{1}{2}\log(z\bar z)\log\left[(1-z)(1-\bar z)\right]\,,
\end{equation}
we find that for all channels, the weight-2 results only contain these two bases $D_2^{\pm}(z)$ as well as their $S_3$ images. 

Now let us look at the singular structure of the weight-2 result. Again we will use the $n_f$ channel result $K$ as the example. Expanding in $z-\bar z\equiv \delta$, we find a very high negative power (obvious from the denominator $(z-\bar z)^{11}$)
\begin{equation}
\label{eq:LT_zmzbexp}
    K\overset{\delta\to 0}{\approx} \frac{1}{\delta^{10}}189 (\bar z-1)^5 \bar z^5 \left(2 \bar z^2-2
   \bar z+1\right) (\bar z \log (1-\bar z)-\log
   (1-\bar z)-\bar z \log (\bar z))+\mathcal{O}(\delta^{-9})\,.
\end{equation}
As we will see in the next section, there should not be any pole in the $z-\bar z\rightarrow 0$ limit, so this condition will strongly constrain the form of the weight-1 coefficient and weight-0 term. We can also look at the squeezed limits $z\rightarrow 0,\, 1,\, \infty$. Take $z\rightarrow 0$ as an example, if we parameterize $z=r\times t, \bar z =r/t$, we get
\begin{equation}
\label{eq:LT_z0exp}
    K\overset{z\to 0}{\approx} \frac{9 t^6 \left(t^8+10 t^6+20 t^4+10 t^2+1\right) (\log
   (r)-1)}{2 r^4 \left(t^2-1\right)^{10}}+\mathcal{O}(r^{-3})\,,
\end{equation}
which contains both power divergence $r^{-4}$ and logarithmic divergence $\log (r)$. The cancellation of these divergences will also highly constrain the lower-weight expressions. Similar behaviors are found for $z\rightarrow 1$ and $z\rightarrow \infty$ squeezed limits.

\subsection{Physical constraints}

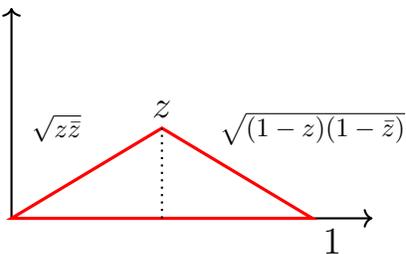
\begin{figure}[!htbp]
\centering
\begin{tikzpicture}[scale=4, thick]
\draw[->, black] (0,0) -- (1.2,0);
\draw[->, black] (0,0) -- (0,0.7);
\draw[red, very thick] (0,0) -- (0.5,0.3) -- (1,0) -- cycle;
\draw[dotted, thick] (0.5,0) -- (0.5,0.3);
\node[above] at (0.5,0.3) {\Large $z$};
\node[below left] at (0,0) {\Large};
\node[below right] at (1,0) {\Large $1$};
\node at (0.15,0.3) {\small $\sqrt{z\bar{z}}$};
\node at (1.0,0.3) {\small $\sqrt{(1 - z)(1 - \bar{z})}$};
\end{tikzpicture}
\caption{The kinematic configuration of the three-point energy correlator in the collinear limit. In the complex $z$ plane, the allowed kinematics forms a triangle with edges $1$, $\sqrt{z\bar z}$ and $\sqrt{(1-z)(1-\bar z)}$.}
\label{fig:e3c_triangle}
\end{figure}

In this subsection, we proceed to the lower-weight terms. Firstly, we need to determine the form of lower-weight ansatz from the singularity structure of the weight-2 functions. The three-point correlator kinematics is described by the conformal ratio variables $z$ and its complex conjugate. Using the definition $x_{13}=x_{12} z\bar z$, $x_{23}=x_{12} (1-z)(1-\bar z)$, as well as the constraint $\tilde{\Delta}_3 <0$, the kinematic space (normalizing $x_{12}=1$) can be represented by a triangle in the complex plane as in Fig.~\ref{fig:e3c_triangle}, with $z$ location of the top vertex. 

\begin{figure}[!htbp]
    \centering
    \includegraphics[scale=0.65]{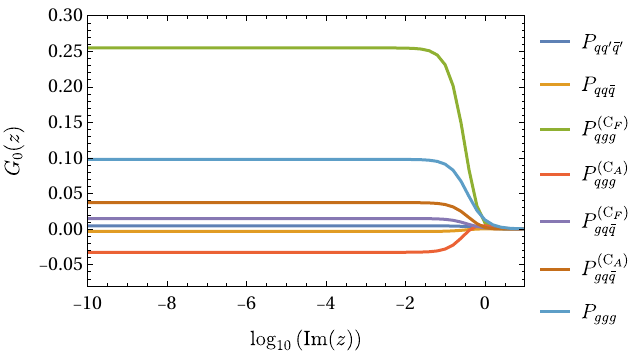}
    \includegraphics[scale=0.69]{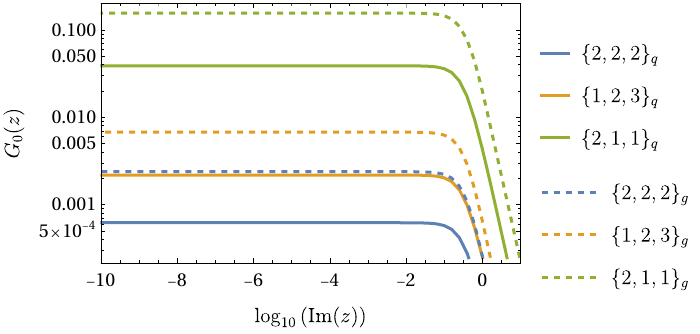}
    \caption{Numerical results for $G_0(z)$ in the collapsed limit. We pick $z=7/11+i 10^x$ and take $x$ from $1$ to $-10$. Left panel: We present the LO EEEC in different partonic channels and show that the collapsed limit is not divergent. Right panel: We show the result with different energy weights: $\{2,2,2\}$, $\{1,2,3\}$ and $\{2,1,1\}$. The subscripts next to the brackets denote quark jet (with $q$, solid lines) and gluon jet (with $g$, dashed lines). All energy weights have no divergence in the collapsed limit.} 
    \label{fig:zmzb_numerics}
\end{figure}

When $z-\bar z\rightarrow 0$, the top vertex approaches the real axis and the triangle collapses into a line. We refer to it as the collapsed limit. It is not a singular limit since it corresponds to a generic hard configuration in the three-point correlator. In Fig.~\ref{fig:zmzb_numerics}, we perform the integration numerically for $\text{Im}(z)=10^x$ and plot $G_0(z=7/11+i 10^x)$: in the left panel, we show all seven partonic channels with $a=b=c=1$; in the right panel, we show the combined quark jet and gluon jet results (labeled in the subscript) with $\{a,b,c\}_{i=q,g}=\{2,2,2\},\,\{1,2,3\},\,\{2,1,1\}$. For all the cases, we confirm that EEEC converges to a constant when $\text{Im}(z)\rightarrow 0$ and thus no divergence.\footnote{For QCD jets, such numerical checks can also be done in the Monte Carlo programs like \textsc{Event2}~\cite{Catani:1996jh,Catani:1996vz}, \textsc{NLOJet++}~\cite{Nagy:2001fj,Nagy:2003tz} and \textsc{EERAD3}~\cite{Gehrmann-DeRidder:2014hxk,Aveleira:2025svg}.} However, as observed in Eq.~\eqref{eq:LT_zmzbexp}, the transcendental weight-2 contribution of the quark $n_f$ channel scales as $1/(z-\bar z)^{10}$ in the collapsed limit and the full expression also involves both weight-1 ($\log(\bar z)$ and $\log(1-\bar z)$) and weight-0 (rational) terms. To guarantee the cancellation of the collapsed divergence, we require that the denominators of weight-1 and weight-0 have the same factor $(z-\bar z)^{n_1}$ and the same power for $n_1$ as the corresponding weight-2 result. Then we need to impose the cancellation of poles in the collapsed limit as the first physical constraint. 
For example, in the quark $n_f$ case, this means that the rational functions in the weight-1 and weight-0 parts should have $1/(z-\bar z)^{10}$.

\begin{figure}
    \centering
    \includegraphics[scale=0.65]{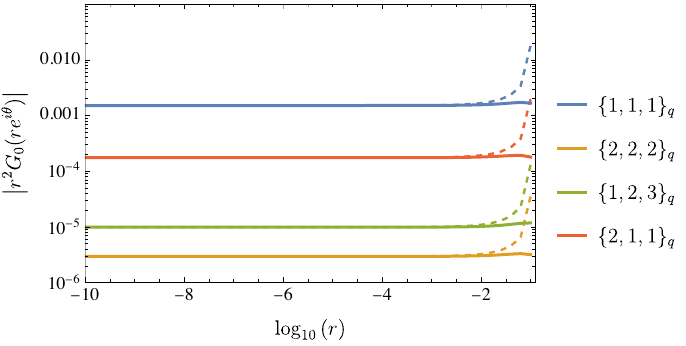}
    \includegraphics[scale=0.65]{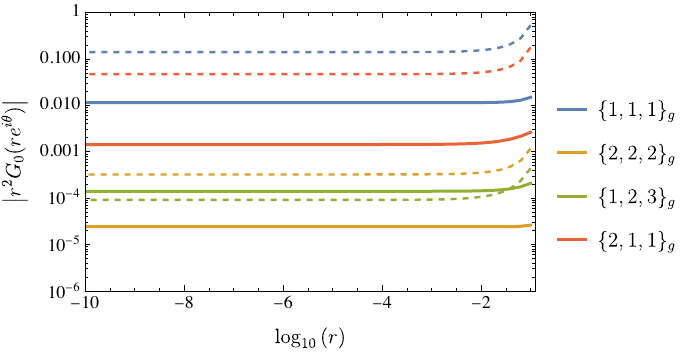}
    \caption{Numerical results for $G_0(z)$ in the squeezed limit $z=0$. With our parameterization, we present the results as $\left|r^2 G_0(r e^{i\theta})\right|$ with respect to $r$. The plateau confirms the $1/r^2$ behavior for all energy weights, which is $1/z\bar z$ in the original parameterization. 
    We pick two values of $\theta$: $\theta=\pi/3$ in the solid lines and $\theta=5\pi/7$ in the dashed lines.
    We show both quark jet (left panel) and gluon jet (right panel).}
    \label{fig:z0_numerics}
\end{figure}

When $z$ approaches $0$, $1$ and $\infty$, we go into the three squeezed regions. Note that the squeezed limit is path-dependent, namely the result can depend on the direction to approach the squeezed pole. To clarify the subtlety, we parameterize this limit via
\begin{align}
\label{eq:r_para}
    z\rightarrow 0:&\qquad z=r\times t,\quad \bar z =\frac{r}{t}\nn\\
    z\rightarrow 1:&\qquad z=1-r\times t,\quad \bar z =1-\frac{r}{t}\nn\\
    z\rightarrow \infty :&\qquad z=\frac{1}{r}\times t,\quad \bar z =\frac{1}{rt}
\end{align}
where $r\rightarrow 0$ will take $z$ to the squeezed limits and $t=\exp(i\theta)$ gives the angular dependence. In Ref.~\cite{Chen:2019bpb}, a factorization formula for leading logarithmic accuracy is derived by sequentially taking the collinear limits of three measured particles, and the resulting ingredients are the two-point energy correlator jet functions with equal energy weight and unequal energy weight. The renormalization group evolutions for these jet functions are both determined by DGLAP kernels~\cite{Dixon:2019uzg}. The authors find that, for example, in the quark jet,
\begin{equation}
    T(x_{12},z,\bar z;a=b=c=1)=\left(\frac{\alpha_s}{4\pi}\right)^2\left[\frac{8}{5}C_F^2+\frac{91}{150}C_F C_A+\frac{13}{300}C_F T_F n_f\right]\frac{1}{x_{12}z\bar z}+\cdots\,,
\end{equation}
with $1/(z\bar z)=1/r^2$ divergence. Note that the squeezed limit originates from the collinear divergence, so there will be at most $1/r^2$ divergence even for $a,b,c>1$. To cross-check, we again perform the numerical integration with $r=10^{-10}$ to $10^{-1}$ with two different values of $\theta=\pi/3$ and $\theta=5\pi/7$ in the squeeze limit $z\rightarrow 0$. Again we look at four choices of energy weights: $\{1,1,1\},\, \{2,2,2\},\, \{1,2,3\},\,\{2,1,1\}$. As shown in Fig.~\ref{fig:z0_numerics}, multiplying the $r^2$ factor, the distributions are almost constant in the small $r$ region, confirming the $1/r^2=1/z\bar z$ behavior. Similarly, we find that the $z\rightarrow 1$ limit scales as $1/r^2=1/(1-z)(1-\bar z)$ and $z\rightarrow\infty$ instead scales as $r^2=1/z\bar z$. To guarantee these asymptotic behaviors, we will put the factor $(z\bar z)^{n_2} ((1-z)(1-\bar z))^{n_3}$ in the denominator, where $n_2,n_3$ are determined by the powers of $z \bar z$ and $(1-z)(1-\bar z)$
in the weight-2 expression and have a maximal value of $1$.
For example, in Eq.~\eqref{eq:LT_z0exp}, the $n_f$ weight-2 contribution scales as $1/r^4$, so we set $n_2=2$ such that the spurious poles $1/r^4$ and $1/r^3$ cancel between weight-2 and lower weights. If the weight-2 contribution does not contain any $1/r$ divergence, we set $n_2=1$, being consistent with the overall asymptotic behavior.
Therefore, the second physical constraint is that $G_0(z)$  scales no worse than $1/r^2$ when $z\rightarrow 0$, as $1/r^2$ when $z\rightarrow 1$ and as $r^2$ when $z\rightarrow \infty$, with the parameterization described above. Any terms like $\ln^m (r)/r^n$ with $n$ higher than expected have to vanish.

Based on the above physical information, we propose the following ansatz:
\begin{align}
\label{eq:ansatz}
    \frac{1}{(z-\bar z)^{n_1} (z\bar z)^{n_2} ((1-z)(1-\bar z))^{n_3}}\Bigg[&\sum_{i,j=0}^{i,j\leq n}a_{i,j} z^i \bar z^j+\sum_{i,j=0}^{i,j\leq n}b_{i,j} z^i \bar z^j\, \log (z\bar z)\nn\\ 
    &+\sum_{i,j=0}^{i,j\leq n}c_{i,j} z^i \bar z^j\, \log ((1-z)(1-\bar z))
    +\sum_{i,j=0}^{i,j\leq n} d_{i,j} z^i \bar z^j \,\pi^2\Bigg]\,.
\end{align}
Here $a_{ij}$, $b_{ij}$, $c_{ij}$ and $d_{ij}$ are unknown parameters.
In the ansatz, we have imposed the single-valued constraint so that the logarithmic arguments are real: $z\bar z$ and $(1-z)(1-\bar z)$. 
The exponent $n_1$ is fixed by the weight-2 expression while the exponent $n_2$ ($n_3$) is determined by the squeezed limit $z\rightarrow 0$ ($z\rightarrow 1$) of the weight-2 result; it has has the maximal value of 1. 
The powers of $z,\bar z$, $i,j$, run from $0$ to $n$, and $n$ is set such that $n-n_1-2n_2-2n_3$ is larger than the highest power of the numerator minus that of the denominator in the weight-2 expression. In practice, one can always increase $n$ and should obtain the same result after imposing all the constraints. Lastly, we add a series of $\pi^2$ terms to correct the weight-2 boundary since we did not fix it when integrating the weight-2 symbols. 

To determine these unknown constants, we impose the following constraints ({\it spurious pole conditions}):
\begin{enumerate}
    \item Any power divergences of $(z-\bar z)$ need to vanish
    \item In the $z\rightarrow 0$ limit, any power divergences or logarithmic divergences that are faster than $\frac{1}{z\bar z}$ (or $\frac{1}{r^2}$) need to vanish. In the meantime, there is no logarithmic divergence with $\frac{1}{z\bar z}$.
    \item Similarly, in the $z\rightarrow 1$ limit, any power divergences or logarithmic divergences that are faster than $\frac{1}{(1-z)(1-\bar z)}$ (or $\frac{1}{r^2}$) need to vanish. In the meantime, there is also no logarithmic divergence with $\frac{1}{(1-z)(1-\bar z)}$.
    \item Lastly, in the $z\rightarrow \infty$ limit, any power divergences or logarithmic divergences that are faster than $\frac{1}{z\bar z}$ (or $r^2$) need to vanish. In the meantime, there is also no logarithmic divergence with $\frac{1}{z\bar z}$.
\end{enumerate}
We will apply the spurious pole conditions sequentially to the sum of weight-2 expressions from the spherical contour and the weight-1+weight-0 ansatz.\footnote{While most of the calculation is done in \textsc{Mathematica}, we perform the series expansion in \textsc{Maple} to speed up the program.} Note that these constraints also determine most of $d_{ij}$, the coefficients of $\pi^2$.

In Tab.~\ref{tab:eeec_111}, we present the number of unknown parameters at each step for $a=b=c=1$. The ansatz column provides the number of initial parameters in the lower-weight ansatz proposed above for $\mathcal{N}=4$ SYM and all QCD channels. The spurious pole conditions ($z-\bar z$, $z=0$, $z=1$ and $z=\infty$) fix more than $90\%$ unknown parameters, and in particular, there are only 6 left for the quark $P_{qgg}^{(\text{C}_F)}$ channel. After that, we perform the so-called {\it symmetrization}, which adds up all six permutations as in Eq.~\eqref{eq:permutations}. We find that some of the terms involving unknown parameters cancel during this step. 

\begin{table}[!htbp]
\centering
\renewcommand{\arraystretch}{1.25}
\begin{tabular}{|c|c|c|c|c|c|c|c|c|c|c|}
\hline
Channel & Ansatz & $z-\bar z$ & $z=1$ & $z=0$ & $z=\infty$ & Symmetrization & $r^{-2}$ & $r^0$ & $r^2$ & $r^4$ \\
\hline
$\mathcal{N}=4$ & 70 & 27  & 26  & 17  & 16  & 12  & 2 & 0  & 0 & 0 \\
\hline
$P_{q q^\prime \bar q^\prime}$ & 664 & 99  & 98  & 97  & 28  & 24  & 13 & 0  & 0 & 0 \\
\hline
$P_{q q\bar q}$ & 931 & 216 & 135 & 134 & 85  & 81  & 59 & 31 & 0 & 0 \\
\hline
$P_{qgg}^{(C_F)}$ & 442 & 27  & 26  & 6   & 6   & 6   & 2  & 0  & 0 & 0 \\
\hline
$P_{qgg}^{(C_A)}$ & 664 & 99  & 98  & 97  & 28  & 24  & 13 & 0  & 0 & 0 \\
\hline
$P_{gq\bar q}^{(C_F)}$ & 511 & 46  & 45  & 21  & 20  & 16  & 5  & 0  & 0 & 0 \\
\hline
$P_{gq\bar q}^{(C_A)}$ & 748 & 133 & 132 & 33  & 32  & 28  & 14 & 0  & 0 & 0 \\
\hline
$P_{ggg}$ & 1134 & 319 & 318 & 197 & 116 & 107 & 86 & 51 & 7 & 0 \\
\hline
QCD Sum                      & 5094 & 939 & 852 & 585 & 315 & 288 & 192 & 82 & 7 & 0 \\
\hline
\end{tabular}
\caption{The number of unknown parameters for $\mathcal{N}=4$ SYM and individual QCD partonic channel at each step. `Ansatz' is the initial lower-weight expression. The next four columns are the spurious pole conditions, and `Symmetrization' denotes adding all permutations. The last four columns correspond to imposing the squeezed limit $z\rightarrow 0$ data and we label them by the power of $r$ in the first equation of Eq.~\eqref{eq:r_para}.}
\label{tab:eeec_111}
\end{table}

Lastly, we need more physical information to fix the remaining parameters. One natural constraint is the leading pole coefficients in the squeezed limit, as we can predict them from factorization. Because we have already summed over all permutations, the behaviors of all three squeezed limits are identical and we only need to look at the $z=0$ limit. In Tab.~\ref{tab:eeec_111}, the $r^{-2}$ shows the undetermined parameters after imposing the leading pole coefficients. Notice that there is more than one parameter fixed at this step, which implies that these unknown parameters are not linearly independent. 

To proceed, we need more squeezed limit data. Another natural direction is to use the power corrections in the $z=0$ limit. In Ref.~\cite{Chen:2021gdk}, the authors are working towards a light-ray OPE formalism to predict these power correction coefficients. However, currently one still needs to power expand the integrand in such limits and perform the phase-space integration (details can be found in the same literature). We adopt the latter approach and evaluate the coefficients for higher powers of $r$. It turns out that once the even-power term in $r$ is fixed (e.g. $r^{-2}$), the next odd-power term is fixed automatically (e.g. $r^{-1}$). This means that we only obtain new information from even-power data. In Tab.~\ref{tab:eeec_111}, we fixed most channels using the next-to-leading power data, while identical quark $P_{qq\bar q}$ and tri-gluon $P_{ggg}$ require $r^2$ and $r^4$ data respectively. We verify that the results we obtain in this way agree with those of Ref.~\cite{Chen:2019bpb}.

\section{Fixing remaining coefficients using lattice reduction}
\label{sec:bootstrap-lattice}

In the bootstrap framework explained in the previous section, we use the power correction data in the squeezed (OPE) limit to determine the remaining parameters after symmetrization. To obtain the data, we still need to perform the analytical integration of the simplified integrand. While it is natural to seek a factorization theorem at subleading powers or a systematic OPE to bypass the explicit calculation, we can also make use of the integrands themselves. Using the analytic regression method with lattice reduction introduced recently~\cite{Barrera:2025uin}, we find that we can reconstruct the remaining parameters directly from the numerical integration data.

In this section, we introduce an alternative approach to bootstrapping the three-point energy correlators, replacing the squeezed limit data with analytic regressions. We will review how lattice reduction works in detail and apply this approach to the computations of three-point correlators with higher energy weights.

\subsection{Lattice reduction}
\label{sec:lattice}

Given the numerical values of an integral and a set of sufficient bases, analytic regression is a method to reconstruct the rational coefficient associated with each basis exactly. In many calculations in quantum field theory and particle physics, people have been using PSLQ~\cite{PSLQref,Bailey:1999nv} widely to rewrite the results in terms of transcendental numbers, like $\pi$ and zeta values $\zeta_n$ (e.g. in the $g-2$ calculation~\cite{Laporta:2017okg}). These transcendental numbers are interpreted as the basis. For complicated examples, PSLQ can require hundreds of digits for numerical evaluations. When the result and basis are functions, in order to perform PSLQ regression, one usually picks a point and turns the function into numbers. However, in Ref.~\cite{Barrera:2025uin}, the authors find that lattice reduction allows for using multiple numeric points in the analytic regression and requires fewer digits than PSLQ.

Given an $d$-dimensional lattice, we can always define it by any set of $d$ independent basis vectors.
Lattice reduction is a method designed to reduce an initial basis into an equivalent basis with the shortest vectors, measured by some norm (e.g. the Euclidean norm of the vector). There are many algorithms for lattice reduction, and the classical one is the Lenstra–Lenstra–Lov\'asz (LLL) algorithm~\cite{Lenstra:1982}~\footnote{There are lots of improved algorithms in the past few decades. Currently both \textsc{Mathematica} and the default choice of C++ program \textsc{FPLLL}~\cite{fplll} use the L${}^2$ algorithm~\cite{Nguyen:2009}.}. 
In the context of Feynman integrals, the lattice basis vectors are constructed using the values of the Feynman integral and the transcendental basis (used to express the Feynman integral) evaluated at various numerical points. Then, lattice reduction is applied to find the integer relations between the Feynman integral and the basis of functions. 

For example, consider the function $f(x)=\frac{1}{3}\text{Li}_2(-x)+\frac{5}{11}\log(x+2)$ as the unknown Feynman integral and $\{B_1(x):=\text{Li}_2(-x), B_2(x):=\log(x+2)\}$ as the set of basis functions which we want to expand $f$ into. 
We evaluate both $f$ and the basis of functions at two points $x_1=2$ and $x_2=5$ keeping 4 digits:
$f(x_1)=0.1512, f(x_2)=-0.03192$ while $B_1(x_1)=\text{Li}_2(-x_1)=-1.437,\, B_1(x_2)=\text{Li}_2(-x_2)=-2.749,\,B_2(x_1)=\log(x_1+2)=1.386,\,B_2(x_2)=\log(x_2+2)=1.946$. 
Then we multiply the number by $10^3$,\footnote{The power of $10$ we multiply all numbers is a parameter of lattice reduction, which is related to the precision digit. Discussions on the choice of this parameter can be found in Ref.~\cite{Barrera:2025uin}.} round into integers and put them into a matrix:
\begin{equation}
    \begin{bmatrix}
        \vec u_1  \\
        \vec u_2 \\
        \vec u_3
    \end{bmatrix}=
    \begin{bmatrix}
        10^3 f(x_1)   & 10^3 f(x_2)   & 1 & 0 & 0 \\
        10^3 B_1(x_1) & 10^3 B_2(x_1) & 0 & 1 & 0 \\
        10^3 B_2(x_2) & 10^3 B_2(x_2) & 0 & 0 & 1
    \end{bmatrix}=
    \begin{bmatrix}
        151 & -32 & 1 & 0 & 0 \\
        -1437 & -2749 & 0 & 1 & 0 \\
        1386 & 1945 & 0 & 0 & 1
    \end{bmatrix}\,,
\end{equation}
where we have also attached the unit vectors into the matrix. Lattice reduction then gives 
\begin{equation}
    \begin{bmatrix}
        \vec v_1  \\
        \vec v_2 \\
        \vec v_3
    \end{bmatrix}=
    \begin{bmatrix}
        0 & 8 & 33 & -11 & -15 \\
        151 & -32 & 1 & 0 & 0 \\
        -25 & 203 & -25 & 8 & 11
    \end{bmatrix}\,,
\end{equation}
Since $\vec u_i$ and $\vec v_i$ span the same lattice, $v_1$ must be a linear combination of $\vec u_i$ with integer coefficients. If we look at the last $3\times 3$ matrix, we learn that $\vec v_1=33 \vec u_1-11 \vec u_2-15\vec u_3$ and thus
\begin{align}
    0&=10^3\left(33f(x_1)-11B_1(x_1)-15B_2(x_1)\right)\nn\\
    8&=10^3\left(33f(x_2)-11B_1(x_2)-15B_2(x_2)\right)
\end{align}
We then conclude that $f(x)=\frac{11}{33}B_1(x)+\frac{15}{33}B_2(x)$. Note that nonzero numbers on the left-hand side of the above equations reflect the fact that we only have finite precision.

The constraints from $r^{-2}$ to $r^4$ in the $z\rightarrow 0$ limit can be replaced by lattice reduction. 
To do so, we define $f$ to be the difference between the true EEEC and the terms that we have determined without using the squeezed limit data. 
We also use the expressions in front of every unknown parameter as the basis of functions (no matter whether they are transcendental or rational).
Consider quark $P_{qgg}^{(\text{C}_F)}$ as an example.
After symmetrization, there are six unfixed coefficients:
\begin{equation}
    G(z) = D(z)+a_{0,11}\,g_1(z)+a_{0,12}\,g_2(z)+a_{1,12}\,g_3(z)+a_{2,12}\,g_4(z)+c_{1,12}\,g_5(z)+d_{2,12}\,g_6(z)\,,
\end{equation}
where $D(z)$ is the part that is completely known and $g_i(z)$ are some $z$-dependent expressions containing both rational functions, $\pi$ and logarithms. For instance,
\begin{align}
    g_3(z)&=\frac{1}{(z-1)^2 z^2 (\bar z-1)^2 \bar z^2}\Bigg[z^4 \bar z^2-z^4 \bar z+8 z^3 \bar z^3-14 z^3 \bar z^2+8 z^3 \bar z-z^3+z^2 \bar z^4-14 z^2 \bar z^3\nn\\
    &+24 z^2 \bar z^2-14 z^2
   \bar z+z^2+\left(2 z^4 \bar z^3-3 z^4 \bar z^2+z^4 \bar z+2 z^3 \bar z^4-8 z^3 \bar z^3+9 z^3 \bar z^2-4 z^3 \bar z+z^3\right.\nn\\
   &-3 z^2
   \bar z^4+9 z^2 \bar z^3-12 z^2 \bar z^2+9 z^2 \bar z-3 z^2+z \bar z^4-4 z \bar z^3+9 z \bar z^2-8 z \bar z+2 z+\bar z^3-3
   \bar z^2\nn\\
   &\left.+2 \bar z\right) \log ((z-1) (\bar z-1))+\left(-2 z^4 \bar z^3+3 z^4 \bar z^2-z^4 \bar z-2 z^3 \bar z^4+8 z^3 \bar z^3-9 z^3
   \bar z^2+2 z^3 \bar z\right.\nn\\
   &\left.+3 z^2 \bar z^4-9 z^2 \bar z^3+6 z^2 \bar z^2-z \bar z^4+2 z \bar z^3\right) \log (z \bar z)-z \bar z^4+8 z
   \bar z^3-14 z \bar z^2+8 z \bar z-\bar z^3+\bar z^2\Bigg]\,.
\end{align}
We will treat $g_i(z)$ as a basis and the unknown parameters as coefficients that require analytic regression. Note that because we simply add all permutations in the symmetrization step and do not remove redundant basis functions, the $g_i(z)$ are not necessarily linearly dependent. 

To proceed, we do the following two steps: (1) run PSLQ or lattice reduction to obtain a linearly independent set of bases (e.g. a linearly independent subset of the $g_{1-6}(z)$ above); 
(2) numerically integrate $G(z)$ to some precision with multiple $z$ values and perform the lattice reduction to determine the parameters. In step (1), we evaluate the bases with some $z$ values and search for linear relations, which can be verified analytically afterward. In practice, we find that even with 1000 digits, PSLQ cannot give all linear relations and thus we use lattice reduction with $100$ values of $z$ and $40-60$ digits. In Tab.~\ref{tab:eeec_111_latt}, we show the number of linearly independent bases after step (1).

In step (2), we need to obtain the numerical values of $G(z)$. As shown in Eq.~\eqref{eq:G0}, we can either perform the two-fold numerical integration or analytically integrate the first one and numerically integrate the second one. The latter approach will allow us to obtain higher precision with less computation cost. With these high-precision numerical data, we can perform the lattice reduction as illustrated in the toy example. The number of points and precisions used for lattice reduction are summarized in Tab.~\ref{tab:eeec_111_latt}. Note that there are tradeoffs between the number of points and digits as discussed in Ref.~\cite{Barrera:2025uin}, but in this table, we do not explore the optimal values. This is because the numerical integration for a single $z $ value only takes a few seconds, and all lattice reductions can be done within minutes. 
We verify the solution against both the one determined from the squeezed limit data and the numerical integration results with values of $z$ that differ from those used for lattice reduction.

\begin{table}[!htbp]
\centering
\renewcommand{\arraystretch}{1}
\begin{tabular}{|c|c|c|c|c|}
\hline
Channel & \# of basis & \# of linearly-independent basis & \# of points & Digits \\
\hline
$\mathcal{N}=4$ &  12 & 6 & 2  & 5  \\
\hline
$P_{q q^\prime \bar q^\prime}$ & 24  & 12 & 10  & 16  \\
\hline
$P_{q q\bar q}$  & 81  & 39 & 30 & 31  \\
\hline
$P_{qgg}^{(C_F)}$ & 6   & 6  & 2  & 13  \\
\hline
$P_{qgg}^{(C_A)}$ & 24  & 12 & 10  & 14  \\
\hline
$P_{gq\bar q}^{(C_F)}$ & 16  & 8  & 3  & 11  \\
\hline
$P_{gq\bar q}^{(C_A)}$ & 28  & 14 & 10  & 14  \\
\hline
$P_{ggg}$ & 107 & 54 & 50 & 22  \\
\hline
\end{tabular}
\caption{A summary of parameter setup in analytic regression with lattice reduction. The first column gives the number of unknown parameters after symmetrization as indicated in Tab.~\ref{tab:eeec_111}, whose coefficients are treated as a set of basis in the lattice reduction. The second column shows the number of independent bases. The last two columns indicate the number of numerical points and the digits we used in the lattice reduction.}
\label{tab:eeec_111_latt}
\end{table}

\subsection{Higher energy weights}
\label{sec:high-energy-weight}

Following a similar procedure, we can also bootstrap the analytic result for higher energy weights, i.e. $a$ or $b$ or $c>1$. As mentioned above, these energy weights are not infrared-safe observables in perturbation theories, though they are measurable in the collider experiments. There are two different ways to make reliable theoretical predictions. One way is attaching the perturbative results to some infrared objects like fragmentation functions or track functions and the infrared poles are removed by renormalization. The other way is to consider instead the projected $N$-point energy correlators~\cite{Chen:2020vvp}, an infrared-safe observable where these higher energy weights are crucial ingredients.

In perturbative theory, projected $N$-point correlators are calculated by integrating the fully differential cross-section, restricting to the largest angular variable. For example, the projected $4$-point correlator at next-to-leading order can be expressed as
\begin{align}
\frac{d\sigma^{[4]}}{dx_L} 
&= \sum_{1 \leq i_1 \neq i_2 \neq i_3 \leq 4} 
\int \mathrm{dLIPS}_4\, |\mathcal{M}_4|^2 
\left(\frac{12 E_{i_1}^2 E_{i_2} E_{i_3}}{Q^3}+\frac{12 E_{i_1} E_{i_2}^2 E_{i_3}}{Q^3}+\frac{12 E_{i_1} E_{i_2} E_{i_3}^2}{Q^3}\right)\nn\\
&\hspace{5cm}\times \delta\left( x_L - \max\{x_{i_1,i_2}, x_{i_1,i_3}, x_{i_2,i_3} \} \right) \nonumber \\
&\quad + \sum_{n \in \{3,4\}} \sum_{1 \leq i_1 \neq i_2 \leq n} 
\int \mathrm{dLIPS}_n\, |\mathcal{M}_n|^2 
\left(\frac{4 E_{i_1}^3 E_{i_2}}{Q^3}+\frac{4 E_{i_1}^3 E_{i_3}}{Q^3}+\frac{4 E_{i_2}^3 E_{i_1}}{Q^3}+\frac{4 E_{i_2}^3 E_{i_3}}{Q^3}\right.\nn\\
&\left.\hspace{2cm}+\frac{4 E_{i_3}^3 E_{i_1}}{Q^3}+\frac{4 E_{i_3}^3 E_{i_2}}{Q^3}+\frac{6 E_{i_1}^2 E_{i_2}^2 }{Q^3}+\frac{6 E_{i_1}^2 E_{i_3}^2 }{Q^3}+\frac{6 E_{i_2}^2 E_{i_3}^2 }{Q^3}\right) \times
\delta\left( x_L - x_{i_1,i_2} \right) \nonumber \\
&\quad + \sum_{n \in \{2,3,4\}} \sum_{1 \leq i_1 \leq n} 
\int \mathrm{dLIPS}_n\, |\mathcal{M}_n|^2 
\frac{E_{i_1}^4}{Q^3} \times 
\delta(x_L)
\end{align}
where the coefficients are from the expansion $(E_{i_1}+E_{i_2}+E_{i_3})^4=E_{i_1}^4+E_{i_2}^4+\cdots$. In order to get the first term, we need the result for the cross-section with energy weights $\{a,b,c\}=\{2,1,1\}, \{1,2,1\}, \{1,1,2\}$. For the second and third terms, our method is not applicable because the integration is divergent in $d=4$. The explicit definition for $N$-point can be found in Ref.~\cite{Chen:2020vvp} and we need all combinations of $a,b,c$ that satisfy $a+b+c=N$.

\begin{table}[!htbp]
\centering
\renewcommand{\arraystretch}{1.25}
\begin{tabular}{|c|c|c|c|c|c|>{\columncolor{blue!15}}c|>{\columncolor{blue!15}}c|}
\hline
Channel & Ansatz & $z-\bar z$ & Squeezed & Sym.  & Indep. & \# of points & Digits \\
\hline
$\mathcal{N}=4$ & 133 & 27  & 16  & 12    &  6  & 3 &  6  \\
\hline
$P_{q q^\prime \bar q^\prime}$ & 931 & 133  & 32  & 28    &  14  & 10 &  16  \\
\hline
$P_{q q\bar q}$ & 1357 & 319 & 122  & 111  & 54 & 30 &  55 \\
\hline
$P_{qgg}^{(C_F)}$ & 585 & 27  & 6  & 6   & 6   & 2 & 13 \\
\hline
$P_{qgg}^{(C_A)}$ & 931 & 133  & 33   & 28  & 14  & 10 &  26  \\
\hline
$P_{gq\bar q}^{(C_F)}$ & 664 & 46  & 20  & 16   &  8 & 3  &   17 \\
\hline
$P_{gq\bar q}^{(C_A)}$ & 931 & 133 & 35  & 28  & 14  & 10 &  17 \\
\hline
$P_{ggg}$ & 1476 & 378 & 141 & 125 & 57 & 50 & 18 \\
\hline
QCD Sum & 6875 & 1169 & 389 & 342 & 167 & - & - \\
\hline
\end{tabular}
\caption{The number of unknown parameters in each step of the bootstrap and the settings of lattice reduction (in blue) for energy weight $a=2,b=1,c=1$. Here `Squeezed' refers to applying the squeezed limit constraints $z=1$, $z=0$ and $z=\infty$ sequentially and `Sym.' stands for symmetrization. In the analytic regression with lattice reduction, we present the number of linearly-independent basis (`Indep.'), the number of $z$ points in the sampling, as well as the digits related to the truncation in the matrix. 
}
\label{tab:eeec_211}
\end{table}

For higher weights, we still follow the same procedure in Fig.~\ref{fig:workflow}. Notice that increasing the values of $a,b,c$ only leads to larger powers in both numerator and denominator after Feynman parameterization, but does not change the quadratic form in the denominator. With the spherical contour method, this leads to the same symbol expressions at weight-2, which means the function spaces for higher energy weights remain the same. In contrast, the residues of each contour, namely the rational functions in front of the transcendental weight-2 functions, will be different. Regarding the physical constraints, as shown in Fig.~\ref{fig:zmzb_numerics} and Fig.~\ref{fig:z0_numerics}, their asymptotic behaviors are the same as $a=b=c=1$, so we also apply the same procedure. The only caution is that $n$, value of the highest power of $z$ and $\bar z$ in our ansatz Eq.~\eqref{eq:ansatz}, can be larger than $a=b=c=1$. To account for it in the program, we start with the smallest possible value of $n$ and if the program does not find a solution, we increase $n$ by $1$ iteratively.

In the following, we show the results for several higher energy weights. Different from the equal energy $a=b=c=1$, we find that symmetrization will regenerate some spurious poles in the squeezed limits and thus we need to apply the constraints again.\footnote{Of course, one can always perform the symmetrization before applying any constraints, but the computation is more expensive.} In Tab.~\ref{tab:eeec_211}, we present the result for $a=2,b=1,c=1$. Instead of using power correction data in the $z\rightarrow 0$ limit, we only use the analytic regression method with lattice reduction discussed in Sec.~\ref{sec:lattice}. We also present the number of independent bases, the number of evaluation points in $z$ as well as the digits we use in the lattice reduction in the table. Again, these values are not optimal, but all reductions are within a reasonable time.

We also perform the same calculation for all possibilities of $a+b+c\leq 5$, which are ingredients for projected energy correlators up to 5-point. In addition, we also evaluate $a=b=c=2$, one term for projected $6$-point energy correlators. Our method can be applied to any values of $a, b, c$ that are greater than or equal to $1$. In Tab.~\ref{tab:eeec_others}, we summarize the number of parameters to be determined at each step for all energy weights. To save space, we no longer present the data for each color-- instead, we only calculate the sum of parameters in all seven QCD channels. After finding the linearly independent basis at the last step, we still perform the lattice reduction for each color channel respectively. In practice, we stick to $40-50$ points for $z$ and evaluate with $50-90$ digits. We perform the reductions with several choices of settings to ensure the analytic regression remains the same (i.e., is correct). After that, we again check the analytic results with the numerical integrations using other choices of $z$.

\begin{table}[!htbp]
\centering
\renewcommand{\arraystretch}{1.25}
\begin{tabular}{|c|c|c|c|c|c|}
\hline
Energy weights & Ansatz & $z-\bar z$ & Squeezed & Sym. & Indep.\\
\hline
\{1,2,1\} & 6984 & 1218 & 395 & 347 & 171  \\
\hline
\{1,1,2\} & 6835 & 1129 & 362 & 318 & 169  \\
\hline
\{2,2,1\} & 9045 & 1478 & 577 & 470 & 214  \\
\hline
\{2,1,2\} & 8871 & 1374 & 555 & 445 & 220  \\
\hline
\{1,2,2\} & 8995 & 1428 & 563 & 471 & 234  \\
\hline
\{3,1,1\} & 9114 & 1477 & 595 & 485 & 235  \\
\hline
\{1,3,1\} & 9357 & 1580 & 611 & 516 & 248  \\
\hline
\{1,1,3\} & 8841 & 1344 & 525 & 430 & 221  \\
\hline
\{2,2,2\} & 11311  & 1703  & 759  & 602 & 281  \\
\hline
\end{tabular}
\caption{The number of unknown parameters for bootstrapping other energy weights: $a+b+c\leq 5$ and $a=b=c=2$. For simplicity, we only provide the sum of all 7 channels in QCD, both quark and gluon jets. The content of the first row is the same as Tab.~\ref{tab:eeec_211}, except that we don't put the number of points and digits used for lattice reduction.}
\label{tab:eeec_others}
\end{table}

\subsection{Results}

\begin{figure}[!htbp]
    \centering
    \includegraphics[width=0.45\linewidth]{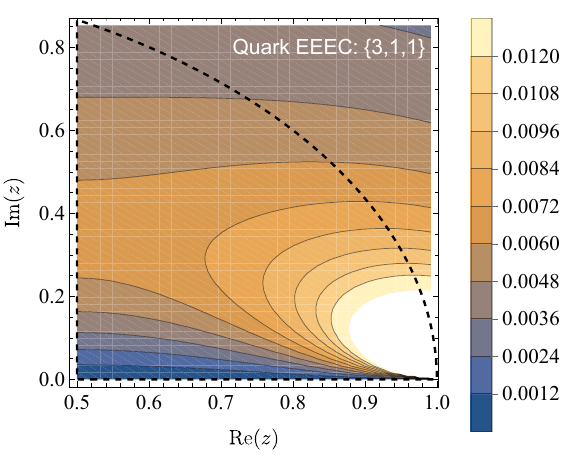}
    \includegraphics[width=0.45\linewidth]{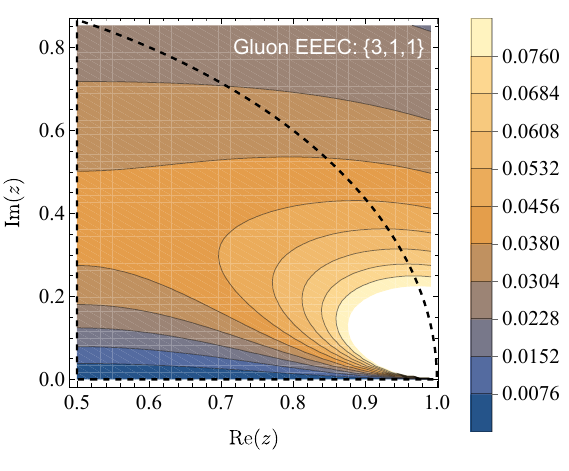}
    \caption{The $G(z)$ distribution with $a=3,b=1,c=1$ in the complex $z$ plane, for both quark and gluon jets. The black dashed lines represent the primary region under $S_3$ symmetry, $1/2\leq\text{Re}(z)\leq 1$ and $0\leq \text{Im}(z)\leq \sqrt{3}/2$. Both quark and gluon share a very similar shape.}
    \label{fig:contourplot}
\end{figure}

Lastly, we present the visualization for some of the analytic results. In Fig.~\ref{fig:contourplot}, we show the $G(z)$ distributions with $a=3,b=1,c=1$ energy weight for both quark jet and gluon jet. Note that with the $S_3$ symmetry from symmetrization, we can identify a primary region in the $z$ plane and obtain the rest region by $S_3$ transformation. In practice, we can choose one angular ordering: $1/2\leq\text{Re}(z)\leq 1$ and $0\leq \text{Im}(z)\leq \sqrt{3}/2$. We indicate the primary region by black dashed lines in the figure. As expected, the distribution gets singular when approaching $z\to 1$ limit and remains regular when approaching $z\to \bar z$ limit. The quark jet and gluon jet distributions have similar shapes; only the absolute values differ. We also find that other energy weights share similar shapes.

\begin{figure}[!htbp]
    \centering    \includegraphics[width=0.75\linewidth]{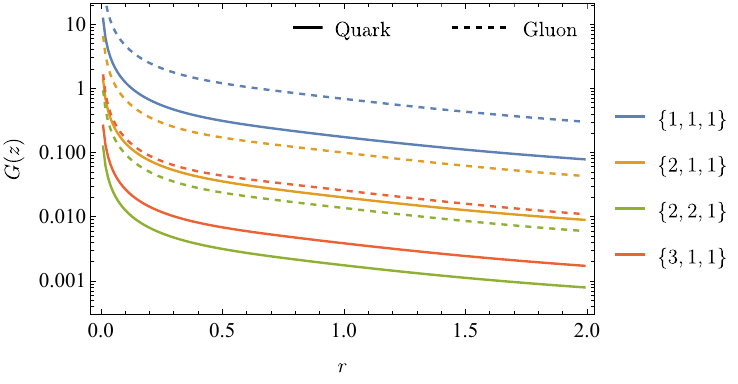}
    \caption{The $G(z)$ distribution with $z=1-re^{i\theta}$ and $\theta=\pi/3$. We present four energy weights $\{a,b,c\}=\{1,1,1\}, \{2,1,1\},\{2,2,1\},\{3,1,1\}$ here for both quark jet (solid) and gluon jet (dashed).}
    \label{fig:rplot}
\end{figure}

\begin{figure}[!htbp]
    \centering    \includegraphics[width=0.75\linewidth]{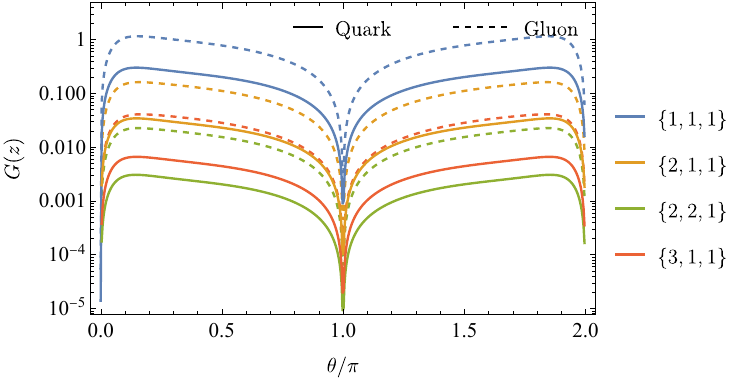}
    \caption{The $G(z)$ distribution with $z=1-re^{i\theta}$ and $r=2/3$. We observe a symmetry $\theta\to 2\pi-\theta$ or equivalently, $z\to \bar z$. The distribution indicates the squeezed limit is direction-dependent.}
    \label{fig:thetaplot}
\end{figure}

To illustrate the singular behavior around the squeeze singularity $z\to 1$, we can again parameterize the kinematics via $z=1-re^{i\theta}$, $\bar z=1-r e^{-i\theta}$. In Fig.~\ref{fig:rplot}, we fix $\theta=\pi/3$ and plot four energy weights in the range $r\in[0,2]$: $\{a,b,c\}=\{1,1,1\}, \{2,1,1\},\{2,2,1\},\{3,1,1\}$. In addition, we also fix $r=2/3$ and explore the angle $\theta$ dependence. As shown in Fig.~\ref{fig:thetaplot}, the distribution is symmetric under $\theta\to 2\pi-\theta$, which is equivalent to $z\to \bar z$. Within $\theta\in[0,\pi]$, $G(z)$ does have a nontrivial dependence on $\theta$. This behavior persists when $r\to 0$, which confirms that the squeezed limit depends on the approaching directions.

In the ancillary file accompanying this paper, we provide the analytic expression of $G(z)$ for $\mathcal{N}=4$ and QCD jets, with all energy weights with $a+b+c\leq 5$ and, in addition, $a=b=c=2$.

\section{Conclusion}
\label{sec:conclusion}
In this work, we initiate an analytic bootstrap framework for three-point energy correlators in QCD and $\mathcal{N}=4$ SYM theory in the homogeneous collinear limit. 
Our approach combines (i) the spherical contour method to determine the complete weight-2 symbol and its rational prefactors 
(ii) stringent physical constraints: single-valued-ness, the absence of spurious poles in the collapsed limit $z\to \bar z$, and the universal scaling in the three squeezed limits $z\to 0,1,\infty$, to fix lower-weight terms, and (iii) an alternative analytic-regression path, using lattice reduction on high-precision numerical data, to determine the remaining coefficients without using additional analytic input. The workflow is summarized in Fig.~\ref{fig:workflow}. While most of the bootstrap literature focuses on $\mathcal{N}=4$ SYM or toy QCD settings, our work has taken a concrete step towards a complete bootstrap program for realistic QCD, which is therefore directly connected to collider measurements and phenomenology.

Using this framework, we derived the leading-order expressions $G(z)$ for $\mathcal{N}=4$ SYM and for all QCD partonic channels, both quark and gluon jets. We find that the transcendental weight-2 function space is universally spanned by the single-valued bases $D_2^{\pm}(z)$ and their $S_3$
images, while spherical contours supply the nontrivial rational functions multiplying these bases.
Particularly, the exact form of these rational functions provides us with some insights into the lower-weight contribution, leading to the ansatz in Eq.~\eqref{eq:ansatz}. We provide a pedagogical derivation for $a=b=c=1$ using both methods, and cross-checked our results with the available data in the literature. We extended the bootstrap beyond equal energy weights and obtained analytic results for all combinations with $a+b+c\leq5$, as well as $\{2,2,2\}$. These higher-weight correlators are necessary ingredients for projected $N$-point energy correlators, which are infrared-safe and of direct phenomenological interest. 

Notably, a key methodological contribution of this work is to apply the analytic regression with the lattice reduction method, bypassing the analytic calculation of power correction data in the squeezed limit. Lattice reduction treats the quantity as a sparse linear combination of some independent basis, no matter whether they are mixed-weight. From high-precision numerical data at various kinematic points, one builds a design matrix and recovers exact rational coefficients by applying lattice basis reduction (e.g. LLL algorithm). In the case of three-point energy correlators, the number of points and digits required for exact regression are both small, leading to a reasonable computational cost. This method can be naturally applied to bypass additional physical constraints or extended to other analytic bootstrap programs.

Our work opens up several directions for future explorations. First of all, it will be interesting to extend the uses of spherical contour algorithm to higher-point energy correlators, other event-shape observables, and general classes of finite integrals. In particular, if the phase-space integrals can be cast into the form of deformed one-loop integrals with a single quadratic denominator, then the problem can be tackled by the spherical contour method. 
Additionally, it will also be interesting to explore if energy correlators share similar mathematical properties to scattering amplitudes (e.g. adjacency relations~\cite{Drummond:2017ssj,Drummond:2018dfd,Caron-Huot:2019bsq}, antipodal duality~\cite{Dixon:2022xqh,Dixon:2023kop,Dixon:2025zwj}, and cluster algebra~\cite{Goncharov:2010jf, Arkani-Hamed:2012zlh,Golden:2013xva,Golden:2014xqa,Chicherin:2020umh,Pokraka:2025ali}).
Secondly, to go beyond single quadratic forms, it will be essential to enhance our method to a full bootstrap program without spherical contours, namely, an analytic QCD bootstrap program. Using the Landau analysis at the integrand level, we can access the alphabet and construct symbol candidates based on integrability. This will allow us to lift the symbol and obtain the raw transcendental function space. The main task is to come up with a robust ansatz for the rational coefficients associated with each transcendental function in the function space, both the leading transcendentality and lower weights. Without the information from the spherical contour, one may need to use a more general denominator: for example, include the entire alphabet in the denominator and search for the minimal power for each factor. Thirdly, as shown in the previous section, the analytic regression method with lattice reduction can save us from tedious calculations in the squeezed (OPE) limits. The precision required for numerical data and computation costs in reduction are both reasonable and affordable. More importantly, lattice reduction can be applied to each order in $\epsilon$, so it can be used for divergent integrals, which will show up in higher-loop orders even for energy correlators. We save these for future work. 

As energy correlators enter a golden era, pushing precision calculations in both $\mathcal{N}=4$ SYM and QCD is essential. Progress will require control over elliptic polylogarithms and beyond: in the collinear limit, one expects elliptic integrals at five-points~\cite{He:2024hbb}; while at generic angles (namely, beyond the collinear limit), elliptic and hyperelliptic sectors are expect at four-points~\cite{Ma:2025qtx}. We also expect more complicated mathematical structures like integrals associated to Calabi-Yau manifolds to appear at higher-loops/higher-points. In this setting, a systematic classification of singularity structures and the associated function spaces of phase-space integrals becomes imperative. A promising avenue is to generalize the projection strategy inspired by spherical contours~\cite{Gong:2025}, while in parallel leveraging high-precision numerical frameworks such as AmRed~\cite{Chen:2024xwt} and AMFlow~\cite{Liu:2017jxz, Liu:2018dmc, Liu:2021wks,Liu:2022chg, Liu:2022mfb}. Coupling these directions with our analytic regression with lattice reduction will provide robust analytic control and scalable exact reconstruction, thereby fostering an analytic bootstrap for energy correlators and making concrete strides toward a QCD bootstrap program.

\acknowledgments

We would like to thank Hofie Hannesdottir and Ellis Ye Yuan for their generous help and
collaboration at the initial stages of this project. 
We also thank Lance Dixon, Jesse Thaler, Hao Chen and Aur\'elien Dersy for many useful discussions.
XYZ is supported by the US Department of Energy under contract DE-SC0013607 and the MIT Pappalardo Fellowship. 
KY and JYG are supported by the National Natural Science Foundation of China (No. 12357077). 
AP was supported in part by the US Department of Energy under contract DE-SC0010010
Task F.

\appendix
\section{$1\to 3$ splitting functions}
\label{app:splitting_function}
For EEEC computations at LO, we need the $1\rightarrow 3$ splitting functions~\cite{Campbell:1997hg,Catani:1998nv,Ritzmann:2014mka}:
\begin{equation}
	\left( \frac{\mu^2 e^{\gamma_E}}{4 \pi} \right)^{2 \epsilon} \frac{4
		g_s^4}{s_{123}^2} P_{1 \rightarrow 3} (z_1, z_2, z_3)
\end{equation}
For $\mathcal{N} = 4$ SYM:
\begin{equation}
	P_{1 \rightarrow 3} (z_1, z_2, z_3) = N_c^2 \left[ \frac{s_{123}^2}{2 s_{13}
		s_{23}} \left( \frac{1}{z_1 z_2} + \frac{1}{(1 - z_1) (1 - z_2)} \right) +
	\frac{s_{123}}{s_{12} z_3} \left( \frac{1}{z_1} + \frac{1}{1 - z_1} \right)
	+ \text{perms} \right]
\end{equation}
For quark jet, we need:
\begin{align}
	P _ { \overline { q } q ^ { \prime } q } =& C _ { F } T _ { F } \frac { s _ { 123 } } { 2 s _ { 12 } } \left[ - \frac { \left[ z _ { 1 } \left( s _ { 12 } + 2 s _ { 23 } \right) - z _ { 2 } \left( s _ { 12 } + 2 s _ { 13 } \right) \right] ^ { 2 } } { \left( z _ { 1 } + z _ { 2 } \right) ^ { 2 } s _ { 12 } s _ { 123 } } + \frac { 4 z _ { 3 } + \left( z _ { 1 } - z _ { 2 } \right) ^ { 2 } } { z _ { 1 } + z _ { 2 } } + ( 1 - 2 \epsilon ) \left( z _ { 1 } + z _ { 2 } - \frac { s _ { 12 } } { s _ { 123 } } \right) \right]\nn\\
	P _ { \overline { q } q q } = & \left( P _ { \overline { q } ^ { \prime } q ^ { \prime } q } + 2 \leftrightarrow 3 \right) + P _ { \overline { q } q q } ^ { ( \mathrm { id } ) }\nn\\
	P _ { \overline { q } q q } ^ { ( \mathrm { id } ) } =&  C _ { F } \left( C _ { F } - \frac { 1 } { 2 } C _ { A } \right) \left\{ ( 1 - \epsilon ) \left( \frac { 2 s _ { 23 } } { s _ { 12 } } - \epsilon \right) + \frac { s _ { 123 } } { s _ { 12 } } \left[ \frac { 1 + z _ { 1 } ^ { 2 } } { 1 - z _ { 2 } } - \frac { 2 z _ { 2 } } { 1 - z _ { 3 } } - \epsilon \left( \frac { \left( 1 - z _ { 3 } \right) ^ { 2 } } { 1 - z _ { 2 } } + 1 + z _ { 1 } - \frac { 2 z _ { 2 } } { 1 - z _ { 3 } } \right) \right. \right.\nn
 \\ &\left. - \epsilon ^ { 2 } \left( 1 - z _ { 3 } \right) ] - \frac { s _ { 123 } ^ { 2 } } { 2 s _ { 12 } s _ { 13 } } z _ { 1 } \left[ \frac { 1 + z _ { 1 } ^ { 2 } } { \left( 1 - z _ { 2 } \right) \left( 1 - z _ { 3 } \right) } - \epsilon \left( 1 + 2 \frac { 1 - z _ { 2 }  } { 1 - z _ { 3 } } \right) - \epsilon ^ { 2 } \right] \right\} + ( 2 \leftrightarrow 3 )\nn\\
	P _ { g g q } =& C _ { F } ^ { 2 } \left\{ \frac { s _ { 123 } ^ { 2 } } { 2 s _ { 13 } s _ { 23 } } z _ { 3 } \left[ \frac { 1 + z _ { 3 } ^ { 2 } } { z _ { 1 } z _ { 2 } } - \epsilon \frac { z _ { 1 } ^ { 2 } + z _ { 2 } ^ { 2 } } { z _ { 1 } z _ { 2 } } - \epsilon ( 1 + \epsilon ) \right] + ( 1 - \epsilon ) \left[ \epsilon - ( 1 - \epsilon ) \frac { s _ { 23 } } { s _ { 13 } } \right] \right.\nn\\
	&\left.+ \frac { s _ { 123 } } { s _ { 13 } } \left[ \frac { z _ { 3 } \left( 1 - z _ { 1 } \right) + \left( 1 - z _ { 2 } \right) ^ { 3 } } { z _ { 1 } z _ { 2 } } - \epsilon \left( z _ { 1 } ^ { 2 } + z _ { 1 } z _ { 2 } + z _ { 2 } ^ { 2 } \right) \frac { 1 - z _ { 2 } } { z _ { 1 } z _ { 2 } } + \epsilon ^ { 2 } \left( 1 + z _ { 3 } \right) \right] \right\}\nn\\
	& + C _ { F } C _ { A } \left\{ ( 1 - \epsilon ) \left( \frac { \left[ z _ { 1 } \left( s _ { 12 } + 2 s _ { 23 } \right) - z _ { 2 } \left( s _ { 12 } + 2 s _ { 13 } \right) \right] ^ { 2 } } { 4 \left( z _ { 1 } + z _ { 2 } \right) ^ { 2 } s _ { 12 } ^ { 2 } } + \frac { 1 } { 4 } - \frac { \epsilon } { 2 } \right) \right.\nn\\
	& + \frac { s _ { 123 } ^ { 2 } } { 2 s _ { 12 } s _ { 13 } } \left[ \frac { 2 z _ { 3 } + ( 1 - \epsilon ) \left( 1 - z _ { 3 } \right) ^ { 2 } } { z _ { 2 } } + \frac { 2 \left( 1 - z _ { 2 } \right) + ( 1 - \epsilon ) z _ { 2 } ^ { 2 } } { 1 - z _ { 3 } } \right]\nn\\
	& - \frac { s _ { 123 } ^ { 2 } } { 4 s _ { 13 } s _ { 23 } } z _ { 3 } \left[ \frac { 2 z _ { 3 } + ( 1 - \epsilon ) \left( 1 - z _ { 3 } \right) ^ { 2 } } { z _ { 1 } z _ { 2 } } + \epsilon ( 1 - \epsilon ) \right]\nn\\
	& + \frac { s _ { 123 } } { 2 s _ { 12 } } \left[ ( 1 - \epsilon ) \frac { z _ { 1 } \left( 2 - 2 z _ { 1 } + z _ { 1 } ^ { 2 } \right) - z _ { 2 } \left( 6 - 6 z _ { 2 } + z _ { 2 } ^ { 2 } \right) } { z _ { 2 } \left( 1 - z _ { 3 } \right) } + 2 \epsilon \frac { z _ { 3 } \left( z _ { 1 } - 2 z _ { 2 } \right) - z _ { 2 } } { z _ { 2 } \left( 1 - z _ { 3 } \right) } \right]\nn\\
	&+ \frac { s _ { 123 } } { 2 s _ { 13 } } \left[ ( 1 - \epsilon ) \frac { \left( 1 - z _ { 2 } \right) ^ { 3 } + z _ { 3 } ^ { 2 } - z _ { 2 } } { z _ { 2 } \left( 1 - z _ { 3 } \right) } - \epsilon \left( \frac { 2 \left( 1 - z _ { 2 } \right) \left( z _ { 2 } - z _ { 3 } \right) } { z _ { 2 } \left( 1 - z _ { 3 } \right) } - z _ { 1 } + z _ { 2 } \right) \right.\nn\\
	&\left.- \frac { z _ { 3 } \left( 1 - z _ { 1 } \right) + \left( 1 - z _ { 2 } \right) ^ { 3 } } { z _ { 1 } z _ { 2 } } + \epsilon \left( 1 - z _ { 2 } \right) \left.\left( \frac { z _ { 1 } ^ { 2 } + z _ { 2 } ^ { 2 } } { z _ { 1 } z _ { 2 } } - \epsilon \right) \right] \right\} + ( 1 \leftrightarrow 2 )
\end{align}
and for gluon jet, we need:
\begin{align}
		P _ { g _ { 1 } q _ { 2 } \overline { q } _ { 3 } } =& C _ { F } T _ { R } P _ { g _ { 1 } q _ { 2 } \overline { q } _ { 3 } } ^ { ( a b ) } + C _ { A } T _ { R } P _ { g _ { 1 } q _ { 2 } \overline { q } _ { 3 } } ^ { ( n a b ) }\nn\\
		P _ { g _ { 1 } q _ { 2 } \overline { q } _ { 3 } } ^ { ( a b ) }  =& - 2 - ( 1 - \epsilon ) s _ { 23 } \left( \frac { 1 } { s _ { 12 } } + \frac { 1 } { s _ { 13 } } \right) + 2 \frac { s _ { 123 } ^ { 2 } } { s _ { 12 } s _ { 13 } } \left( 1 + z _ { 1 } ^ { 2 } - \frac { z _ { 1 } + 2 z _ { 2 } z _ { 3 } } { 1 - \epsilon } \right) \nn\\ 
		& - \frac { s _ { 123 } } { s _ { 12 } } \left( 1 + 2 z _ { 1 } + \epsilon - 2 \frac { z _ { 1 } + z _ { 2 } } { 1 - \epsilon } \right) - \frac { s _ { 123 } } { s _ { 13 } } \left( 1 + 2 z _ { 1 } + \epsilon - 2 \frac { z _ { 1 } + z _ { 3 } } { 1 - \epsilon } \right)\nn\\
		P _ { g _ { 1 } q _ { 2 } \overline { q } _ { 3 } } ^ { ( n a b ) } =& \left\{ - \frac { t _ { 23,1 } ^ { 2 } } { 4 s _ { 23 } ^ { 2 } } + \frac { s _ { 123 } ^ { 2 } } { 2 s _ { 13 } s _ { 23 } } z _ { 3 } \left[ \frac { \left( 1 - z _ { 1 } \right) ^ { 3 } - z _ { 1 } ^ { 3 } } { z _ { 1 } \left( 1 - z _ { 1 } \right) } - \frac { 2 z _ { 3 } \left( 1 - z _ { 3 } - 2 z _ { 1 } z _ { 2 } \right) } { ( 1 - \epsilon ) z _ { 1 } \left( 1 - z _ { 1 } \right) } \right] \right.\nn\\ 
		& + \frac { s _ { 123 } } { 2 s _ { 13 } } \left( 1 - z _ { 2 } \right) \left[ 1 + \frac { 1 } { z _ { 1 } \left( 1 - z _ { 1 } \right) } - \frac { 2 z _ { 2 } \left( 1 - z _ { 2 } \right) } { ( 1 - \epsilon ) z _ { 1 } \left( 1 - z _ { 1 } \right) } \right] \nn\\
		& + \frac { s _ { 123 } } { 2 s _ { 23 } } \left[ \frac { 1 + z _ { 1 } ^ { 3 } } { z _ { 1 } \left( 1 - z _ { 1 } \right) } + \frac { z _ { 1 } \left( z _ { 3 } - z _ { 2 } \right) ^ { 2 } - 2 z _ { 2 } z _ { 3 } \left( 1 + z _ { 1 } \right) } { ( 1 - \epsilon ) z _ { 1 } \left( 1 - z _ { 1 } \right) } \right] \nn\\ 
		& \left.- \frac { 1 } { 4 } + \frac { \epsilon } { 2 } - \frac { s _ { 123 } ^ { 2 } } { 2 s _ { 12 } s _ { 13 } } \left( 1 + z _ { 1 } ^ { 2 } - \frac { z _ { 1 } + 2 z _ { 2 } z _ { 3 } } { 1 - \epsilon } \right) \right\} + ( 2 \leftrightarrow 3 )\nn\\
		P _ { g _ { 1 } g _ { 2 } g _ { 3 } } =& C _ { A } ^ { 2 } \left\{ \frac { 1 - \epsilon } { 4 s _ { 12 } ^ { 2 } } t _ { 12,3 } ^ { 2 } + \frac { 3 } { 4 } ( 1 - \epsilon ) + \frac { s _ { 123 } } { s _ { 12 } } \left[ 4 \frac { z _ { 1 } z _ { 2 } - 1 } { 1 - z _ { 3 } } + \frac { z _ { 1 } z _ { 2 } - 2 } { z _ { 3 } } + \frac { 3 } { 2 } + \frac { 5 } { 2 } z _ { 3 } \right.\right.\nn\\ 
		& \left.+ \frac { \left( 1 - z _ { 3 } \left( 1 - z _ { 3 } \right) \right) ^ { 2 } } { z _ { 3 } z _ { 1 } \left( 1 - z _ { 1 } \right) } \right] + \frac { s _ { 123 } ^ { 2 } } { s _ { 12 } s _ { 13 } } \left[ \frac { z _ { 1 } z _ { 2 } \left( 1 - z _ { 2 } \right) \left( 1 - 2 z _ { 3 } \right) } { z _ { 3 } \left( 1 - z _ { 3 } \right) } + z _ { 2 } z _ { 3 } - 2 + \frac { z _ { 1 } \left( 1 + 2 z _ { 1 } \right) } { 2 } \right.\nn\\
		& \left.\left.+ \frac { 1 + 2 z _ { 1 } \left( 1 + z _ { 1 } \right) } { 2 \left( 1 - z _ { 2 } \right) \left( 1 - z _ { 3 } \right) } + \frac { 1 - 2 z _ { 1 } \left( 1 - z _ { 1 } \right) } { 2 z _ { 2 } z _ { 3 } } \right] \right\} + ( 5 \text { permutations} )
\end{align}
where
\begin{equation*}
	t _ { i j , k } = 2 \frac { z _ { i } s _ { j k } - z _ { j } s _ { i k } } { z _ { i } + z _ { j } } + \frac { z _ { i } - z _ { j } } { z _ { i } + z _ { j } } s _ { i j }.
\end{equation*}

\section{Geometric interpretation}\label{sec:geometryintepretion}
In this appendix, we simply introduce the geometric interpretation of both first entry conditions and the spherical contour $S^2$. More detailed discussion can be found in Ref.~\cite{Arkani-Hamed:2017ahv, YelleshpurSrikant:2019khx,Gong:2025}.
\subsection{First Entry} The first entries $s_{n_1}$ are related to branch points~\cite{aomoto_1982,Duhr:2012fh,Duhr:2014woa,Weinzierl:2022eaz}
\begin{equation}
    s_{n_1}=0 \quad \text{and} \quad s_{n_1}=\infty.
\end{equation}
That means when we set $s_1=0$ (or $1/s_1=0$), the geometric configurations between the integral contours and the singularity hypersurfaces will become singular. We use $\mathcal{Q}$ to denote the singularity hypersurface $\mathcal{Q}:XQX=q_{ii}x_i^2+2q_{ij}x_ix_j+q_{jj}x_j^2+\mathrm{rest.}=0$ and $\overline{V_iV_j}$ to denote the $\mathbb{CP}^1$ through the two vertices $V_i$ and $V_j$. Note the remaining terms in "$\mathrm{rest.}$" do not contain the crossing term $x_ix_j$ and thus do not affect the geometry between $\mathcal{Q}$ and $\overline{V_iV_j}$. We will focus on the former part. Let us get back to the first entry conditions \eqref{eq:firstentry}.
\begin{enumerate}
    \item [(a)] If $q_{ii}\neq0\,\,\text{and}\,\,q_{jj}\neq0$, the first entries can be separated into 
    \begin{equation}
        \otimes\,r(Q^{(ij)})=\otimes {(q_{ij}-\sqrt{q_{ij}^2-q_{ii}q_{jj}})\over q_{ii}}-\otimes{(q_{ij}+\sqrt{q_{ij}^2-q_{ii}q_{jj}})\over q_{ii}}.
    \end{equation}
    We add a factor $1/q_{ii}$ by hand so that the first entries seem like the solution of the equation in the affine patch
    \begin{equation}
        q_{ii}x_i^2+2q_{ij}x_ix_j+q_{jj}x_j^2|_{x_j=1}=0.
    \end{equation}
    That means, when setting the first entry to $0$ to find the branch points, the vertex $V_i$ will be located on $\mathcal{Q}$. Similarly, if we add $1/q_{jj}$, $V_j$ will be on $\mathcal{Q}$. Therefore, first entry conditions are vertices on the singularity hypersurfaces.

    \item [(b)] If $q_{ii}=0$ and $q_{jj}\neq0$ (or the same as exchanging $i,j$), the quadric will become $2q_{ij}x_ix_j+q_{jj}x_j^2$, and $V_i$ has already been on $\mathcal{Q}$. First entries can be separated into two parts
    \begin{equation}
        \otimes \left(\frac{q_{ij}^{2}}{q_{ii}}\right)^{-2\text{sign}(q_{ij})}=-2\mathrm{sign}(q_{ij})\left(\otimes q_{ij}^2-\otimes q_{ii}\right).
    \end{equation}
    The former is related to a branch point $q_{ij}=0$. When we reach this branch point, $\overline{V_iV_j}$ is tangent to $\mathcal{Q}$ at the point $V_i$. The prefactor $2$ means we can get back to the result when deforming $q_{ij}$ around the branch twice. The latter one is related to a branch point $q_{jj}=0$, and this is to say $V_{j}$ is also on $\mathcal{Q}$.

    \item [(c)] If $q_{ii}=0$ and $q_{jj}=0$, both of the vertices $V_i$ and $V_j$ are on the quadric. The first entry is now $q_{ij}^{-2\mathrm{sign}(q_{ij})}$. The branch point is $q_{ij}=0$. When we reach the branch point, the $\mathbb{CP}^1:\overline{V_iV_j}$ is completely embedded in the quadric $\mathcal{Q}$.
\end{enumerate}
To conclude, the first entries are the expression that when they are set to $0$ or $\infty$, the geometric configuration of singularity hypersurfaces and integral contours will become more singular.

\subsection{Spherical Contour} There are two aspects when performing the spherical contour to calculate discontinuities: why spherical contour works, and what the remaining quadric $Q^{(ij)}$ is. We will give a simple introduction to both of the problems.

    \paragraph{Projection quadric $Q^{(ij)}$}: The new quadric singularities $Q^{(ij)}$ in Eq.~\eqref{eq:projQij} can be viewed as a line projection of $Q$ from $\overline{V_iV_j}$. We can express it in another description. Assume that the vertices of the integral contour except for $V_i$ and $V_j$ span a codim-2 hyperplane $H_{\{\widehat{{i,j}}\}}$ (or to say the $\mathbb{CP}^{n-3}$ space of discontinuity integrals), where the quadric $Q^{(ij)}$ lives. Any point $P\in Q^{(ij)}$ along with $V_i$, $V_j$ forms a two-dimensional plane $H_{(ijP)}$. This plane $H_{(ijP)}$ is tangent to the original $Q$ (only one intersection point).
    \begin{equation}
        Q^{(ij)}:\{P\in H_{\{\widehat{{i,j}}\}}|\, H_{(ijP)} \,\text{is tangent to}\, Q\}.
    \end{equation}
    Algebraically, if we parametrize points on $H_{(ijP)}$ by\footnote{Any point in $\mathbb{CP}^{n-1}$ can be represented by $n$ independent points because of the projective property.} 
    \begin{equation}
        V=P+\alpha_iV_i+\alpha_jV_j,
    \end{equation}
    the tangent condition can be written as
    \begin{equation}
        VQV=\partial_{\alpha_i}(VQV)=\partial_{\alpha_j}(VQV)=0.
    \end{equation}
    The solutions of $P$ will form the new quadric $Q^{(ij)}$.

\begin{figure}[ht]
\centering
\begin{tikzpicture}[scale=0.8, line cap=round, line join=round]
  \fill (4.25,4.2) node[right] {$\mathbb{CP}^n$};
  \draw[thick] (4.25,4.5) -- (4.25,3.8) -- (5.2,3.8);
  \coordinate (Vi) at (-3.2, 3.6);
  \coordinate (Vj) at ( 2.8, 4.2);
  \draw[thick] (Vi) -- (Vj);
  \fill (Vi) circle (2pt) node[left]  {$V_i$};
  \fill (Vj) circle (2pt) node[right] {$V_j$};
  \coordinate (G1) at (-3-1,-0.9);
  \coordinate (G2) at (3.4+0.3+0.3+1,-0.9);
  \coordinate (G4) at (-4-1,-3.1);
  \coordinate (G3) at (2.4+0.3+0.3+1,-3.1);
  \draw[fill=gray!30!white] (G1) -- (G2) -- (G3)-- (G4) -- cycle;
  \coordinate (O) at (0,1);
  \draw[thick] (O) circle (2); 
  \coordinate (Q) at (1.5,3); 
  \fill (Q) node[left] {$Q$};
  \fill (4.25,-0.9) node[left,below] {\scriptsize $\mathbb{CP}^{n-2}$};
  \coordinate (Tc) at (0,-2);  
  \def\rx{1.6} \def\ry{0.8};
  \draw[thick] (Tc) ellipse (\rx cm and \ry cm);
  \foreach \ang in {200,0} {
    \pgfmathsetmacro{\xx}{\rx*cos(\ang)}
    \pgfmathsetmacro{\yy}{\ry*sin(\ang)}
    \coordinate (P\ang) at ($(Tc)+(\xx,\yy)$);
    \fill (P\ang) circle (1.6pt);
    \draw (Vi) -- (P\ang);
    \draw (Vj) -- (P\ang);
    \pgfmathsetmacro{\dx}{0.08*\ry*sin(\ang)}
    \pgfmathsetmacro{\dy}{0.08*\rx*cos(\ang)}
  }
  \pgfmathsetmacro{\xxx}{\rx*cos(200)}
  \pgfmathsetmacro{\yyy}{\ry*sin(200)}
  \pgfmathsetmacro{\xxxx}{\rx*cos(0)}
  \pgfmathsetmacro{\yyyy}{\ry*sin(0)}
  \coordinate (V) at (-1.2,3.8);
  \coordinate (Vp) at (0.4,3.96);
  \fill ($(Tc)+(\xxx,\yyy)$) node[left] {\scriptsize $P$};
  \fill ($(Tc)+(\xxxx,\yyyy)$) node[right] {\scriptsize $P^\prime$};
  \fill (1.6,-2.5) node[below=0.5cm,right=-0.1cm] {$Q^{(ij)}$};
  \draw[dotted] ($(Tc)+(\xxx,\yyy)$) -- (V);
  \draw[dotted] ($(Tc)+(\xxxx,\yyyy)$) -- (Vp);
  \coordinate (P) at ($(Tc)+(\xxx,\yyy)$);
  \coordinate (M) at ($(V)!0.5!(P)$);
  \coordinate (Pp) at ($(Tc)+(\xxxx,\yyyy)$);
  \coordinate (N) at ($(Vp)!0.47!(Pp)$);
  \fill (M) circle (1.6pt) node[left] {\scriptsize $V$};
  \fill (N) circle (1.6pt) node[right] {\scriptsize $V^\prime$};
\end{tikzpicture}
\caption{$Q^{(ij)}$ is the projection of the quadric $Q$ through $\overline{V_iV_j}$. Here $V$ and $V^\prime$ indicate the tangent points of plane $H_{(ijP)}$ and $H_{(ijP^\prime)}$ to the quadric $Q$.}
\end{figure}
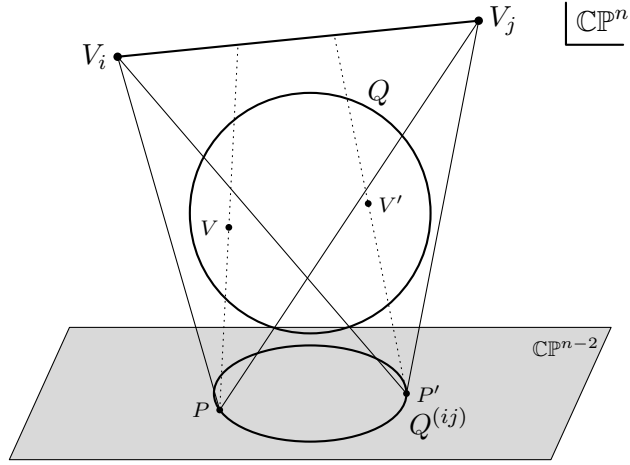

\paragraph{$S^2$ spherical contour}: The spherical contour $S^2$ is a two-dimensional integral contour. In the algorithm we introduced previously, we use the parametrization $r,\theta$ to perform the integration. However, it can be decomposed into two $S^1$ contours, separately related to $x_i$ and $x_j$, and the order of the contours does not affect the discontinuities. The $S^1$ contour is defined similar to the spherical contour by performing $\{x_i\}\mapsto\{w_i\}$ so that the quadric takes the form
    \begin{equation}
        x_i^2+X_{\widehat{\{i\}}}Q^{(i)}X_{\widehat{\{i\}}},
    \end{equation}
    and integrate $x_i$ from $-\infty$ to $\infty$. However, two $S^1$ contours only reduce one transcendental weight. This implies that the two $S^1$ contours play different roles: One for discontinuities which reduces the weight by $1$, and one for a geometric contour identity. It is worth noting again that the leading transcendental weight of an iterated integral with a quardic singularity is $n/2$ for even $n$ and $(n-1)/2$ for odd $n$. This indicates the two $S^1$ contours exchange their functions between even and odd $n$. In other words, if we perform a $S^1$ contour to an iterated integral, this $S^1$ either calculates discontinuities or indicates the contour identity.
    
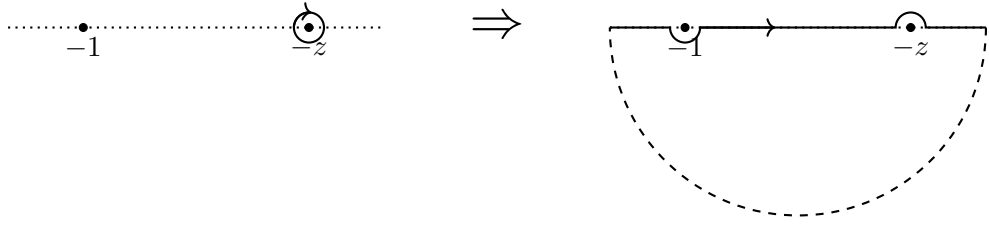
\begin{figure}[h]
    \begin{center}
    \begin{tikzpicture}
    \begin{scope}[xshift=-4cm]
    \draw [black,thick,dotted] (0,0) -- (5,0);
    \coordinate [label=-90:$-z$] (u2) at (4,0);
    \coordinate [label=-90:$-1$] (u3) at (1,0);
    \filldraw (u2) circle [radius=1.5pt];
    \filldraw (u3) circle [radius=1.5pt];
    \draw [thick] (u2) circle [radius=.2];
    \draw [thick,->] ($(u2)-(.2,0)$) arc [start angle=180, end angle=80,radius=.2];
    \end{scope}
    \begin{scope}[xshift=4cm]
    \draw [black,thick,dotted] (0,0) -- (5,0);
    \coordinate [label=-90:$-z$] (v2) at (4,0);
    \coordinate [label=-90:$-1$] (v3) at (1,0);
    \filldraw (v2) circle [radius=1.5pt];
    \filldraw (v3) circle [radius=1.5pt];
    \draw [thick,dashed] (5,0) arc [start angle=0,end angle=-180, radius=2.5];
    \draw [thick] (0,0) -- ($(v3)-(.2,0)$) arc [start angle=180, end angle=360, radius=.2] -- ($(v2)-(.2,0)$) arc [start angle=180, end angle=0, radius=.2] -- (5,0);
    \draw [thick,->] ($(v3)+(.2,0)$) -- +(1,0);
    \end{scope}
    \node [anchor=center] at (2.5,0) {\huge $\Rightarrow$};
    \end{tikzpicture}
    \caption{Residue and $S^1$ contour of log example in affine space.}
    \label{fig:firstS1example}
    \end{center}
    \end{figure}

    Let us take even $n$ cases in \eqref{eq:Idef} as an example to introduce the contour for discontinuities. The discontinuities of an iterated integral are computed by taking residues, which can be deformed to an $S^1$ contour. This is visualized by the logarithm example. Following the first integral in \eqref{eq:example}, the logarithm can also be written in the projective space with a quadric singularity
    \begin{equation}
        \log(z)=\int_{\Delta_2}\frac{(z-1)\la X\d X\ra}{(x_1+z x_2)(x_1+x_2)}=\int_{\Delta_2}\frac{(z-1)\la X\d X\ra}{(x_1^2+(1+z)x_1x_2+zx_2^2)}.
    \end{equation}
    Its discontinuity can be calculated by taking the residue around $x_1=-zx_2$. This contour can be deformed along the real axis and the lower half-plane circle, which yields an $S^1$ contour (see Fig.~\ref{fig:firstS1example}).

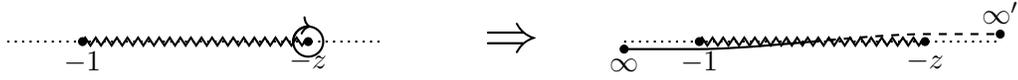
\begin{figure}[h]
    \begin{center}
    \hspace{0.5cm}
    \begin{tikzpicture}
    \begin{scope}[xshift=-4cm]
    \draw [black,thick,dotted] (0,0) -- (5,0);
    \coordinate [label=-90:$-z$] (u2) at (4,0);
    \coordinate [label=-90:$-1$] (u3) at (1,0);
    \filldraw (u2) circle [radius=1.5pt];
    \filldraw (u3) circle [radius=1.5pt];
    \draw [thick] (u2) circle [radius=.2];
    \draw [thick,->] ($(u2)-(.2,0)$) arc [start angle=180, end angle=80,radius=.2];
    \draw [thick,decoration={zigzag,segment length=4pt, amplitude=1.5pt},decorate] (u3) -- (u2);
    \end{scope}
    \hspace{0.2cm}
    \begin{scope}[xshift=4cm]
    \draw [black,thick,dotted] (0,0) -- (5,0);
    \coordinate (u1) at (3,0);
    \coordinate [label=-90:$-z$] (u2) at (4,0);
    \coordinate [label=-90:$-1$] (u3) at (1,0);
    \coordinate [label=-90:$\infty$] (inf1) at (0,-.1);
    \coordinate [label=90:$\infty'$] (inf2) at (5,.1);
    \filldraw (u2) circle [radius=1.5pt];
    \filldraw (u3) circle [radius=1.5pt];
    \draw [thick] (inf1) -- ($(u3)+(0,-.1)$) .. controls ($(u3)+(.5,-.1)$) and (2.5,0) .. (2.5,0);
    \draw [thick,dashed] (2.5,0) .. controls (2.5,0) and ($(u2)+(-.5,.1)$) .. ($(u2)+(0,.1)$) -- (inf2);
    \filldraw  (inf1) circle [radius=1.5pt];
    \filldraw  (inf2) circle [radius=1.5pt];
    \draw [thick,decoration={zigzag,segment length=4pt, amplitude=1.5pt},decorate] (u3) -- (u2);
    \end{scope}
    \node [anchor=center] at (2.5,0) {\huge $\Rightarrow$};
    \end{tikzpicture}
    \caption{$S^1$ contour in the cases with branch cuts.}
    \label{fig:firstS1generic}
    \end{center}
    \end{figure}

    In more general cases, $n+k$ in the integral definition \eqref{eq:Idef} can be odd. The geometry of this $S^1$ contour is slightly different from the even $n+k$ case. The solution of $XQX=0$ now gives two branch points instead of poles. Therefore, the "residue" contour is torn apart into two pieces by the branch cut. It will deform into a curve instead of an actual $S^1$ contour, whose ends are located at infinity in different Riemann sheets. An example is shown in Fig. \ref{fig:firstS1generic}, in which the integrand has branch points at $x=-1$ and $x=-z$. Note that this example integral yields a different algebraic function from the logarithmic function.

    The other $S^1$ contour does not contribute to the discontinuities, but is related to a contour identity. We first note that if we perform an $S^1$ contour to an integral and it does not give a discontinuity, the integrand is a total derivative, or to say a differentiation of a lower weight form that is algebraic. Its corresponding contour is still a lower-dimensional standard simplex. A typical example is $\mathrm{Li}_2(z)$. Recall that the $\mathrm{Li}_2(z)$ is defined in \eqref{eq:li2}, and it is an integration over a total derivative
    \begin{equation}
        \mathrm{Li}_2(z)=\frac{1}{4}\int_{\Delta_5} \d\left(\frac{2\left(\la X\d X^3\ra_{(1234)}+\la X\d X^3\ra_{(1345)}+z\la X\d X^3\ra_{(1245)}-\la X\d X^3\ra_{(2345)}\right)}{(x_5^2+x_5((1-z)x_1+x_2+x_3+x_4)+x_1x_3+x_1x_4+x_2x_4)^2}\right)
    \end{equation}
    where $\la X\d X^3\ra_{(ijkl)}$ indicates the measure in the complex space $[x_i:x_j:x_k:x_l]\in \mathbb{CP}^3$. For generic geometric configurations that no vertices of the standard simplex are on the singularity hypersurface (i.e., all the $x_i^2$ exist), therefore, one can localize the integrand to the closed boundary by Stokes' theorem to calculate the integral. The canonical simplex contour then can be viewed as a sum over all possible $S^1$ contours with respect to different variables (up to a factor related to the dimension since the simplex has been repeatedly counted). If we denote the integral after a contour identity $S^1$ corresponding to $x_i$ by $I_{n,k}^{(i)}$, the contour identity $S^1$ means
    \begin{equation}\label{eq:georelation}
        \mathcal{S}I_{n,k}=\frac{1}{n-1}\sum_{i=1}^n\mathcal{S}I_{n,k}^{(i)}.
    \end{equation}
    For example, if we consider a general scalar integral in $\mathbb{CP}^2$
    \begin{equation}
        I_{\mathrm{ex}}=\int_{\Delta_{2}}\frac{\sqrt{a_{12}^2+a_{13}^2+a_{23}^2-2a_{12}a_{23}a_{13}-1}\la X\d X^2\ra}{2\sqrt{2}(x_1^2+x_2^2+x_3^2+a_{12}x_1x_2+a_{13}x_1x_3+a_{23}x_2x_3)^{\frac{3}{2}}}.
    \end{equation}
    The factor is the determinant of the quadric with an opposite sign, which we use $q_{123}$ to denote. Its symbol then from first entry conditions is
    \begin{equation}\label{eq:symbolgeoex}
        \mathcal{S}I_{\mathrm{ex}}=\frac{1}{2}\left(\otimes \frac{-2a_{12}+a_{13}a_{23}-4q_{123}}{-2a_{12}+a_{13}a_{23}-4q_{123}}+\mathrm{sym.}\right).
    \end{equation}
    One can easily check the sum of all $S^1$ contours is
    \begin{equation}
        \sum_{i=1}^3\int_{\Delta_{(i)}}\la X_{\widehat{\{i\}}}\d X_{\widehat{\{i\}}}\ra\int_0^\infty\frac{q_{123}\d x_i}{(x_i^2+X_{\widehat{\{i\}}}Q^{(i)}X_{\widehat{\{i\}}})^{\frac{3}{2}}}=\log\left(\frac{-2a_{12}+a_{13}a_{23}-4q_{123}}{-2a_{12}+a_{13}a_{23}-4q_{123}}\right)+\mathrm{perm},
    \end{equation}
    which yields twice of the symbol \eqref{eq:symbolgeoex}. This illustrates the relation \eqref{eq:georelation} in $\mathbb{CP}^2$. For cases that some vertices are on the singularity, the story would be much harder and remain for further understanding. But the spherical contour method to calculate symbols still works.

\bibliographystyle{JHEP}
\bibliography{refs.bib}

\end{document}